\documentclass{jfm}
\usepackage{amsmath}
\usepackage[dvipsnames]{xcolor}
\usepackage[squaren]{SIunits}

\begin{document}

\newtheorem{lemma}{Lemma}
\newtheorem{corollary}{Corollary}

\title{Statistical steady state in turbulent droplet condensation}
\author[C. Siewert, J. Bec, and G. Krstulovic]{Christoph Siewert,\ns J{\'e}r{\'e}mie Bec, and Giorgio Krstulovic}

\shorttitle{Statistical steady state in turbulent droplet condensation} 


\affiliation{Universit\'e C\^ote d'Azur, Observatoire de la C\^ote
  d'Azur, CNRS, Laboratoire Lagrange, 06300 Nice, France.}

\maketitle

\begin{abstract}
  Motivated by 
  systems in which droplets grow and shrink in a turbulence-driven
  supersaturation field, we investigate the problem of turbulent
  condensation in a general manner. Using direct numerical
    simulations we show that the turbulent fluctuations of the
  supersaturation field offer different conditions for the growth of
  droplets which evolve in time due to turbulent transport and
  mixing. Based on that, we propose a Lagrangian stochastic model for
  condensation and evaporation of small droplets in turbulent
  flows. It consists of a set of stochastic integro-differential
  equations for the joint evolution of the squared radius and the
  supersaturation along the droplet trajectories. The model has two
  parameters fixed by the total amount of water and the thermodynamic
  properties, as well as the Lagrangian integral timescale of the
  turbulent supersaturation.  The model reproduces very well the
  droplet size distributions obtained from direct numerical
  simulations and their time evolution. A noticeable result is that,
  after a stage where the squared radius simply diffuses, the system
  converges exponentially fast to a statistical steady state
  independent of the initial conditions.  The main mechanism involved
  in this convergence is a loss of memory induced by a significant
  number of droplets undergoing a complete evaporation before growing
  again.  The statistical steady state is characterised by an
  exponential tail in the droplet mass distribution. These results
  reconcile those of earlier numerical studies, once these various
  regimes are considered.
\end{abstract}

\section{Introduction}
\label{sec:intro}

There are many systems where it is necessary to quantify the
rate at which droplets shrink and grow in a turbulent environment.
The efficiency and environmental impact of a number of
  propulsion or energy production systems (diesel engine, gasoline
  engines with direct injection, cryogenic rocket engines, steam
  turbines, fuel cell \dots) depend on the control of evaporation or
  condensation processes.  In natural contexts the prime example is
  the radiative transfer due to clouds on Earth \citep{gw13} and other
  planets \citep{ingersoll2004dynamics}, which strongly depends upon
  the microphysics of the droplet growth.  In both engineered and
  natural situations turbulence can have a strong influence \citep[see
  for example][]{reveillon2007effects, db12}.  Due to the variety of
  such systems we want to reevaluate the problem of turbulent
  condensational growth in a general manner and ask, from a fluid
  dynamicist viewpoint, what are the effects of supersaturation
  fluctuations and of the turbulent transport of droplets on the
  evolution of the droplet size distribution.  Therefore, we want to
  find a simplified description of the general problem, which still
  covers the main physical process, identify the main parameters and
  study the basic behavior of the system as a function of these
  parameters.

Most work on droplet condensation in turbulence is motivated
  and applied to terrestrial clouds.  An open question is whether
  turbulent effects can bridge the droplet size gap in which neither
  the classical growth by diffusion, nor the classic
  collision-coalescence growth are efficient \citep{s03,db12,gw13}.  A
  condensational broadening of the droplet size distribution by
  turbulent transport and mixing would increase the collision
  likelihood.  However, using mean-field arguments
  \citet{bartlett1972dispersion} came to the conclusion that the
  droplet size distribution stays narrow also when turbulent updrafts
  are taken into account.  Such mean-field arguments have anyhow two
  drawbacks: Firstly, as noticed by \citet{srivastava1989growth}, the
local supersaturation in the direct vicinity of a droplet can differ
from one droplet to another and from the average
supersaturation. These differences in local supersaturation can be due
to variations in the amount of vapor consumption by droplets with
different sizes as well as fluctuations due to turbulence. Indeed, it
was found using stochastic modeling
\citep{kulmala1997effect,khvorostyanov1999toward, mcgraw2006brownian}
that, due to such fluctuations, droplets can exists even if the global
supersaturation is negative.  Hence, in such models the droplet size
spectra are broader than those obtained by mean-field models.

The second drawback of mean-field models is the assumption that the
tracked volume does not mix during the evolution.  
However, it is well known that this is generally not true in
turbulent flows. For example, inertialess droplets (tracers) separate
explosively following Richardson diffusion, leading to a strong mixing. 
Based on this picture
\citet{lasher2005broadening} and \citet{sidin2009lagrangian} used
large-eddy simulations and kinematic simulations in which they
backtrace all droplets that are in a specific volume at a final time.
As they come from very different locations, a broad droplet size
distribution is observed in the final volume.
However, such a two-step approach
is not well suited to account for the mutual influence and competition
between droplets that were close to each other in the past.

One way to overcome these issues is to perform
direct numerical simulations (DNS) for the joint evolutions of the
fluid velocity, temperature, and water vapor, without introducing any ad-hoc
modelling. This approach naturally accounts for the above-mentioned
points, the spatial and temporal fluctuations in the supersaturation
and the turbulent transport and mixing. However, such simulations are
computationally very demanding and became thus feasible only rather
recently.  \citet{vaillancourt2002microscopic} were the first to
perform DNS of the turbulent fields coupled with a Lagrangian droplet
model.  They find only a very little increase in the broadening of the
droplet size spectra, compared to simulations without turbulence
\citep{vaillancourt2001microscopic} and explain this by the
decorrelating effect of turbulence between the droplet size and the
supersaturation in its vicinity.

\citet{celani2005droplet} have performed two-dimensional DNS in order
to increase the domain size and the range of turbulent eddies.  They
find the opposite trend, namely a very broad droplet size spectrum,
but which might originate from neglecting the influence of the
droplets on the supersaturation field.  Later they contrast their
results incorporating more realistic physical treatments.  In
\citet{celani2008equivalent}, they implement a detailed droplet
activation scheme and obtain that the droplet size spectra are still
very broad, independently of the activation process.  In
\citet{celani2007droplet} and \citet{celani2009droplet}, they account
for the back reaction of droplets on the supersaturation field and
observe a negative mean supersaturation and a reduction of the
spectral broadening that eventually becomes stationary.  Nevertheless,
the spectra they measure are still very broad.

\citet{paoli2009turbulent} have performed three-dimensional DNS of
turbulent condensation with a large-scale forcing acting not only on
the fluid velocity but also on the temperature and vapor fields.
However, they do not specifically focus on estimating spectral
broadening but rather on developing a stochastic model intended to be
used in large-eddy or Reynolds-averaged simulations.
\citet{lanotte2009cloud} find in 3D DNS that the droplet surface area
has a distribution very close to a Gaussian and that the broadening is
rather small, in agreement with the findings of
\citet{vaillancourt2002microscopic}.  However this broadening,
measured for instance by the standard deviation of the droplet surface
area, increases with both time and Reynolds number.  By dimensional
arguments, they extrapolate this behaviour to the large Reynolds
numbers of clouds and predict a significant broadening.  Very
recently, \citet{sardina2015continuous} repeated the investigations of
\citet{lanotte2009cloud} with longer simulation times and higher
resolutions.  They found that the standard deviation of the droplet
surface area increases proportionally to the square root of time.
Additionally, they make use of a stochastic model very similar to the one of
\citet{paoli2009turbulent} to obtain that the constant factor in this
behaviour is proportional to the Reynolds number.  

Such an approach is particularly promising for
designing realistic models for droplet condensation in turbulent
flows. However, the various contradicting results of DNS described
above still lack a clear understanding.  They show either a strong
broadening or very few broadening when the results
are not extrapolated to high Reynolds numbers. Questions
remain on the origins of these differences: Are they due to
dissimilarities in settings or governing equations? Does the problem
lack of universality and depend on dimensionality, initial conditions,
and activation processes?

The present work aims at providing some answers to these
  questions. The paper is organised as follows: In
  \S\ref{subsec:GoverningEqs} we first describe the governing
  equations of our simplified system.  The variations in temperature
  and vapor concentration are modeled in terms of a single scalar
  field, the supersaturation, which is passively advected by the flow.
  Droplets are passively transported by a homogeneous, isotropic
  turbulent flow.  The droplets can completely evaporate, without
  disappearing, and grow again if they reach a
  positive-supersaturation region.  Based on dimensional analysis
  (\S\ref{subsec:HandWavyingArguments}), the turbulent phase change is
  described by two relevant timescales: the droplet growth rate and
  the response time of the supersaturation field upon condensation or
  evaporation. Depending on whether these characteristic times are
  greater or smaller than turbulent timescales and the observation
  time, one expects different regimes.  We confirm the existence of
  these regimes by three-dimensional direct numerical simulations
  (\S\ref{subsec:DNS}).  We then introduce
  in~\S\ref{subsec:ModelDerivation} a stochastic integro-differential
  Lagrangian model, which imposes the global mass conservation of
  liquid and vapor and is expected to be valid for timescales much
  larger than the Lagrangian correlation time.
  In~\S\ref{subsec:ModelDNSComparison} this model is shown to
  reproduce well the results of DNS.  In addition we find that the
  condensation process converges to a statistically steady state, so
  that the droplet size distribution and the fluctuations of the
  supersaturation field become independent of time and initial
  conditions.  For both the model and the DNS, the stationary
  probability distribution function (PDF) of droplet masses is shown
  to have an exponential tail (\S\ref{subsec:TheoResults} and
  \S\ref{subsec:SteadyStateComparison}) that we characterise in
  \S\ref{subsec:SteadyStateCharacterization} as a function of the
  relevant parameters.  We show that after a Brownian stage during
  which the droplet surface area just diffuses
  (\S\ref{subsec:ShortTimeBehavior}), the exponential convergence to
  the steady state occurs once a significant fraction of droplets has
  completely evaporated at least once
  (\S\ref{subsec:LongTimeBehavior}).  Finally, we draw concluding
  remarks in \S\ref{sec:conclusion}.  A list of symbols can be found
  in Appendix~\ref{appendixA}.

\section{General Framework}
\label{sec:settings}

To concentrate on the influence of turbulence we choose idealized conditions. 
This enables us to make analytical predictions but, of
course, at the risk of missing important physical phenomena that might
be important for applications. That is why we state precisely in the
following under which conditions the equations are valid.
The equations are stated in a general form in
  \S\ref{subsec:GoverningEqs}, the numerical treatment will be
  described in \S\ref{subsec:DNS}.

\subsection{Governing Equations}
\label{subsec:GoverningEqs}

The immediate neighborhood of a given droplet is characterized
  by a local value of the supersaturation field $s = p/p_s-1$ (where
  $p$ is the vapour partial pressure and $p_s$ the saturation
  pressure), which induces the growth of its squared radius $r^2$,
  i.e. the droplet surface area:
\begin{equation}
  \label{eq:SurfaceGrowth}
\frac{dr^2}{dt} = \left\{ \begin{array}{ll} 2 a_3 s &  r^2 \geq  0, \\
                            0 & r^2 = 0 \; \& \; s < 0. \end{array}
                        \right.
\end{equation}
The coefficient $a_3$ is assumed constant, \textit{i.e.}\/ the small
temperature dependence is neglected.  It is here understood that the
timescale associated with the establishment of vapour diffusion is
faster than the external timescales especially the smallest scales of
turbulence, the Kolmogorov scale $\eta$ \citep[see,
\textit{e.g.},][for a detailed derivation]{pk97b,
  vaillancourt2001microscopic}.  This approximation might become
invalid in some highly turbulent environments encountered in technical
applications where $\eta$ can be very small.
Additionally, it is assumed that the droplet volume loading is small
enough to ensure no overlap between the diffusion regions of different
droplets.
Finally, curvature and salinity effects relevant for the activation
and affecting the growth of very small droplets are neglected
\citep{celani2008equivalent}.  Instead, we assume that a completely
evaporated droplet is simply reactivated if it is located in a region
with positive supersaturation playing the role of a condensation
nuclei that stays in the system.  We show in \S\ref{sec:SteadyState}
that the specific type of boundary condition imposed at $r=0$ is not
important for the large-value tail of the droplet size distribution,
\textit{i.e.}\/ faraway from the zero-size boundary.

The growing and shrinking droplets are transported by a turbulent
gas velocity $\mathbf{u}$ according to the Stokes drag law
\begin{equation}
  \label{eq:DropletVelocity}
  \frac{d\mathbf{v}}{dt} = -\frac{1}{\tau_d\left(r\right)}
  \left[\mathbf{v}-\mathbf{u}(\mathbf{x},t)\right],
\end{equation}
where $\mathbf{x}$ and $\mathbf{v}$ are the particle position and
velocity, $\tau_d = 2 \rho_d r^2 / (9 \rho \nu)$ is the droplet
response time, $\rho_d$ its mass density, and $\rho$ and $\nu$ are the
gas mass density and kinematic viscosity, respectively.  The Stokes
drag approximation is valid for very small and heavy liquid droplets.
The droplet radius $r$ has to be smaller than the smallest scale of
turbulence, the Kolmogorov dissipative scale
$\eta = \nu^{3/4}/\varepsilon^{1/4}$, where $\varepsilon$ is the mean
kinetic energy dissipation rate of the turbulent gasflow.  Although
the liquid to gas density ratio $\rho_d/\rho$ is typically large, the droplet size and therewith its mass are typically
small enough for gravitational acceleration to be neglected.  A
posteriori we know that in our parameter range the droplet motion is
practically identical to tracer motion as the droplet response time
stays negligibly small.  Furthermore, for the smallest droplets
Brownian motion might become important.

The gas velocity $\mathbf{u}$ evolves according to the
incompressible Navier-Stokes equation
\begin{equation}\label{eq:FluidVelocity}
  \frac{\partial \mathbf{u}}{\partial t} + \mathbf{u}\cdot\nabla
  \mathbf{u} = - \frac{1}{\rho} \nabla p + \nu \nabla^2 \mathbf{u} +
  \phi_u \,, \quad \nabla\cdot\mathbf{u}=0\,,
\end{equation}
where $p$ is the pressure and $\phi_u$ is a large-scale
  forcing, which maintains turbulence in a developed regime.  Mixing
  with the environment at the boundaries, often called entrainment and
  detrainment, is neglected.  Additionally, as for the droplets, we
  neglect temperature and gravity effects, including buoyancy.  This
  might be the strongest approximation as latent heat release upon
  condensation is often a strong source for natural moist convection.
  However, without the anisotropies introduced by buoyancy and edge
  effects, we generate a homogeneous and isotropic turbulent flow.
  This drastically simplifies the statistics and their interpretation
  and thus motivated us to use these approximations as a first step.
As a matter of fact, we will show that some of our
results are similar to those of \citet{celani2007droplet} who
accounted for the temperature field and buoyancy effects.

Since thermal effects are neglected, it is not necessary to treat the
temperature and vapour fields separately.  Instead the supersaturation
$s$ can be modeled as passive scalar coupled to the Lagrangian
particles
\begin{equation}
  \label{eq:SupersaturationField}
  \frac{\partial s}{\partial t} + \mathbf{u}\cdot\nabla s = \kappa
  \nabla^2 s - \sum_{i=1}^N 4 \pi \rho_d a_2 a_3 r_i\, s(\mathbf{x}_i,t)\,
  \delta\!\left(\mathbf{x}_i-\mathbf{x}\right) + \phi_s \,,
\end{equation}
where $\kappa$ is the molecular diffusivity of the vapour inside the
gas.  The second term on the right-hand side accounts for the local
change in supersaturation due to the presence of $N$ droplets.  There,
$\mathbf{x}_i$ denotes the position of the $i$th droplet and the
coefficient $a_2$ accounts for both the change in vapour mass and
temperature due to the condensation or evaporation.  We assume again
that the temperature dependence of $a_2$ is negligible. There
  are two common choices for the forcing term $\phi_s$.  The first
  one, often used in experiments on passive scalars, is to impose a
  mean gradient \citep{warhaft2000passive}.  In atmospheric conditions
  there is often a vertical temperature gradient, so that updrafts
  increase the supersaturation
  \citep{squires1952growth,twomey1959nuclei}.  Hence, the DNS studies
  in the context of cloud physics (see \S\ref{sec:intro}) considered
  $\phi_s = \mathbf{u} \cdot (0, 0, a_1)$, where $a_1$ is a constant
  that depends weakly on temperature.  If $a_1$ is assumed to be
  constant, this forcing is just a special form of the mean gradient
  forcing.  The second choice, more commonly used in the physics
  community and in numerical simulations on the advection of passive
  scalars by a turbulent flow, is to force also the supersaturation
  field by a large-scale forcing.  The properties of the
  supersaturation field without droplets, which is then a passive
  scalar, are thus well known.  Even if this has been recently
  questioned by~\citet{PhysRevLett.115.114502}, one expects such
  properties to not depend on the specific form of forcing and to
  display some universality.  For instance, coherent structures are
  present at all length scales, separated by regions where the field
  undergoes significant fluctuations, the so-called fronts associated
  with very sharp gradients \citep{celani2001fronts}. 
  If there is no mean forcing input ($\int_\upsilon\phi_s d^3 x= 0$ ), the mean value of supersaturation remains constant.
  Then, the system (\ref{eq:SurfaceGrowth})-(\ref{eq:SupersaturationField})
  conserves the global mass of liquid and vapour
\begin{equation}
  w \upsilon = \int_\upsilon \left( s(x,t)+1 \right)\,\mathrm{d}^3 x + a_2\sum_{i=1}^N \rho_d\frac{4}{3}\pi\,r_i(t)^3 = \mathrm{const} \, .
  \label{eq:conser_total_mass}
\end{equation}
Similar mass balances have been proven to be useful to obtain
  analytical results
  \citep{pinsky2013supersaturation,devenish2016analytical}.
Motivated by our idea to concentrate on the influence of
  turbulent mixing, we choose here to use a large-scale forcing for
  $\phi_s$.  We will show in \S\ref{sec:conclusion} that we can
  explain the results of those DNS studies with updraft forcing by the
  outcomes of our simulations.  However, in contrast to assuming a
  mean gradient, our choice ensures the statistical homogeneity and
  isotropy of the scalar field $s$, and hence of the droplet sizes.

\subsection{Qualitative Predictions}\label{subsec:HandWavyingArguments}
In the various systems (see \S\ref{sec:intro}) the physical
  parameters such as the kinematic viscosity of the gas, the number of
  droplets and the total amount of liquid and vapour can take very
  different values.  To treat the problem of turbulent condensation as
  generally as possible, \textit{i.e.}\/ independently of any specific
  application, we analyse here dimensionless equations.  Thereby,
  non-dimensional groups are obtained by dimensional analysis that are
  determining the solutions of the system of differential equations.
  As long as the assumptions made in the previous subsection are not
  violated, the solutions we obtain can be applied to a specific
  application, once rescaling them together with time and lengthscales
  using the specific dimensional parameters. Non-dimensional
  quantities are always written in capital letters to distinguish them
  from the dimensional quantities written in lowercase.

The turbulent fluctuations of supersaturation along droplet
trajectories are essentially correlated over the large scales, because
to leading order, droplets are Lagrangian objects. We thus decide to
use as a reference length, the large scale
$l_0=u_{\rm rms}^3/\varepsilon$, where $u_{\rm rms}$ is the
root-mean-square velocity, and as a reference time the large eddy
turnover time $t_0=l_0/u_{\rm rms}$.

Applying this to the
Navier-Stokes equations (\ref{eq:FluidVelocity}) leads to the
Reynolds number
\begin{equation}\label{eq:ReynoldsNumber}
Re = \frac{\tau_\nu}{t_0} = \frac{l_0^2/\nu}{t_0}\,,
\end{equation}
which is the ratio of viscous mixing time scale to the turbulent
advection time scale. In principle we want the Reynolds number as
large as possible in order to have a large separation between the
forcing and the dissipation scales.  However, the computational effort
increases with the resolution, such that most of our simulations will
be conducted at $Re = 2100$.

The droplet equation of motion (\ref{eq:DropletVelocity}) leads to
introduce the large-scale Stokes number
\begin{equation}\label{eq:StokesNumber}
  St = \frac{\tau_d}{t_0} = \frac{ 2 \rho_d \bar{r}^2}{9 \rho \nu t_0}.
\end{equation}
A priori the droplet sizes evolve with time, and so
do the Stokes numbers. Motivated by the global mass balance eq.~(\ref{eq:conser_total_mass})
we choose here and in the following to use the
droplet mass averaged over the whole population as typical size $\bar{r} = f(t)$, namely
$\bar{r} = \langle r(t)^3\rangle^{1/3}$.  As the droplets are typically
small, we expect the Stokes number to be small, so that inertia
effects, such as clustering, are weak even on the smallest scales.

The supersaturation equation~(\ref{eq:SupersaturationField}) provides
two timescale ratios: The Schmidt number
\begin{equation}\label{eq:PrandtlNumber}
Sc = \frac{\tau_\kappa}{\tau_\nu} = \frac{l_0^2/\kappa}{l_0^2/\nu},
\end{equation}
that we only consider to be one and the timescale of change in
supersaturation due to condensation
\begin{equation}\label{eq:SupersaturationTimeScale}
\frac{\tau_s}{t_0} =  \frac{1}{4 \pi \rho_d a_2 a_3 n_d \bar{r}
  t_0}, 
\end{equation}
which is obtained from a spatial average and depends on both 
the characteristic radius $\bar{r}$ and the droplet
number density $n_d$.

The supersaturation diffusion-advection
equation~(\ref{eq:SupersaturationField}) is linear and hence the
proper choice for the reference value of $s$ is not straightforward.
We select its standard deviation $s_{\rm rms}$ in the absence of droplets.
From the droplet growth equation (\ref{eq:SurfaceGrowth}) a time scale
arises, which we call hereafter the condensation timescale
\begin{equation}
  \label{eq:CondensationTimeScale}
  \frac{\tau_c}{t_0} =  \frac{\bar{r}^2}{2 a_3 s_{\rm rms} t_0}.
\end{equation}
This condensation timescale did not receive much attention in recent
DNS studies of homogeneous condensation
\citep{vaillancourt2002microscopic,celani2005droplet,lanotte2009cloud,sardina2015continuous}. However
it is known and used in the context of mixing of sub- and supersaturated regions.
Based on the notation in reacting flows a Damk\"ohler number is
constructed to compare the turbulent mixing timescale to the time for
phase change \citep{db12}.  In this sense, the dimensionless
quantities defined in Eqs.~(\ref{eq:SupersaturationTimeScale}) and
(\ref{eq:CondensationTimeScale}) can be seen as inverse Damk\"ohler
numbers.  This raises the question about the relevant timescales that
characterise the system.  From their DNS results,
\citet{kumar2012extreme} concluded that the Damk\"ohler number based
on the supersaturation timescale is the relevant one.  However,
\citet{lehmann2009homogeneous} argued that the timescale of the
coupled system might be very different from both the condensation and
supersaturation timescales.

We give here some preliminary heuristic arguments on possible
asymptotic behaviour of the system. Let us consider the supersaturation
equation~(\ref{eq:SupersaturationField}): If $\tau_s$ is very large,
the influence of the particle on the supersaturation field is small,
\textit{i.e.}\/ its evolution is dominated by turbulent mixing
characterised by the timescale $t_0$.  Conversely, if this
supersaturation timescale is very small, the condensation or
evaporation of droplets dominates the variations of $s$ and the
characteristic timescale for the supersaturation evolution is
$\tau_s$. The supersaturation field thus changes on a timescale given
by the minimum of $t_0$ and $\tau_s$.  Assume now that the
condensation timescale $\tau_c$ is much larger than the minimum of
$t_0$ and $\tau_s$, meaning that the droplet sizes change slowly
compared to the supersaturation field.  This was implicitly assumed in
mean-field models and corresponds to the homogeneous case
(small Damk\"ohler number) in the context of chemical reactions.  In
this case individual droplets experience a fast changing
supersaturation, which to leading order can be approximated as a
random white noise.  From Eq.~(\ref{eq:SurfaceGrowth}), one infers
that their surface area $r^2$ follows a Brownian motion.  This implies
that the standard deviation of the droplets surface $\sigma_{r^2}$,
which gives a measure of the spectral broadening, evolves as
$\propto\sqrt{t}$.  Conversely, if we now consider that $\tau_c$ is
much smaller than the minimum of $t_0$ and $\tau_s$, the droplet sizes
change faster than the supersaturation value along their
trajectories. We are then in the inhomogeneous limit, where more
complex dynamics can be expected.  Droplets experience a quasi
constant supersaturation value for a long time and change their sizes
rapidly.  This could lead to both very large droplets and complete
evaporation.  We thus expect in this case a very broad droplet size
spectrum.

\subsection{Direct Numerical Simulations}
\label{subsec:DNS}

\begin{table}
\begin{center}
    \begin{tabular}{rrrrrrrrrrrr}
\multicolumn{1}{c}{$N_x^3$} & \multicolumn{1}{c}{$\frac{\nu}{l_*^2/t_*}$} & \multicolumn{1}{c}{$\frac{\eta}{l_*}$} & \multicolumn{1}{c}{$\frac{\tau_\eta}{t_*}$} & \multicolumn{1}{c}{ $\frac{u_\mathrm{rms}}{l_*/t_*}$}   & \multicolumn{1}{c}{$\frac{l}{l_*}$} & \multicolumn{1}{c}{$\frac{t_0}{t_*}$} & \multicolumn{1}{c}{$\frac{t_{u}}{t_*}$} & \multicolumn{1}{c}{$\frac{t_{s}}{t_*}$}  & \multicolumn{1}{c}{$Re$} & \multicolumn{1}{c}{$R_\lambda$} \\\hline 
$256^3$  & $3.1\,10^{-3}$ & $1.32\,10^{-2}$ & $5.7\,10^{-2}$  & 1.20  & 1.83 & 1.53 & 0.31 & 0.72 &   715 & 100 \\ \hline    
$512^3$  & $4.6\,10^{-4}$ & $3.1\,10^{-3}$ & $2.2\,10^{-2}$ &  1.0 & 1.0 & 1.0 & 0.36 & 0.64 & 2175 & 180 \\ \hline
$1024^3$ & $1.53\,10^{-4}$ & $1.4\,10^{-3}$ & $1.2\,10^{-2}$ & 1.04 & 1.03 & 1.02 & 0.32 & 0.60 & 6800 & 320    
    \end{tabular}
  \end{center}
  \caption{\label{table1}
    Parameters of the numerical simulations for the turbulent gasflow with respect to the large-scale values of the intermediate resolution case denoted as $l_*$ and $t_*$:
    $N_x^3$ number of spatial collocation points;
    $\nu$ kinematic viscosity; 
    $\eta = \nu^{3/4}/\varepsilon^{1/4}$ Kolmogorov dissipative scale;
    $\tau_\eta = \nu^{1/2}/\varepsilon^{1/2}$ Kolmogorov time;
    $u_\mathrm{rms}$ root-mean square velocity;
    $l_0 = u_\mathrm{rms}^3/\varepsilon$ large scale;
    $t_0= u_\mathrm{rms}^2/\varepsilon$ large-eddy turnover time;
    $t_{u}$ integral timescale of the Lagrangian velocity autocorrelation;
    $t_{s}$ integral timescale of the Lagrangian scalar autocorrelation;
    $Re = u_{rms} l_0/\nu$ large-scale Reynolds number;
    $R_\lambda = \sqrt{15}\, u^2_\mathrm{rms} \tau^2_\eta / \eta^2$ Taylor microscale Reynolds number.} 
\end{table}

We conduct a series of direct numerical simulations of the flow and
the supersaturation field with the pseudo-spectral code LaTu
\citep{homann2007impact} in a cubic, periodic domain. The
  gas flow forcing $\phi_u$ is chosen to keep constant the energy
  content of the first two shells of wavenumbers in Fourier space.
  The supersaturation forcing $\phi_s$ is Gaussian, white-noise in
  time and concentrated at wavenumbers with moduli between $1$ and
  $2.5$. The various non-dimensional parameters are
reported in Table~\ref{table1}.  Most simulations are performed at a
resolution of $512^3$ collocation points, corresponding to a Taylor
micro-scale Reynolds number $R_\lambda \approx 180$.  The ratio
between the largest and the smallest time scales is
$t_0/\tau_\eta \approx 50$, indicating a sufficient scale separation.
Additionally, selected simulations are repeated at decreased ($256^3$)
and increased ($1024^3$) resolutions, corresponding to
$R_\lambda\approx100$ and $R_\lambda\approx320$, respectively, in
order to test possible Reynolds number dependence.  We only consider
cases where the Schmidt number is unity and the Eulerian mean
supersaturation is initially zero. The fields are evolved until a
statistical steady state is reached. Then ten millions equal-sized
droplets are randomly released in the domain with a large-scale Stokes
number of $St = 2.3\times 10^{-5}$ to ensure that particle
  inertia is initially negligible.  These Lagrangian droplets,
together with their sizes, are subsequently integrated according to
Eqs.~(\ref{eq:SurfaceGrowth}) and (\ref{eq:DropletVelocity}) using a
tri-cubic interpolation of the gas velocity and of the supersaturation
value at the particle position.  The back-reaction of the droplets
onto the supersaturation field in Eq.~(\ref{eq:SupersaturationField})
is performed by linear extrapolation to the nearest collocation
points.

\begin{figure}
  \centerline{ \includegraphics[width=0.65\textwidth]{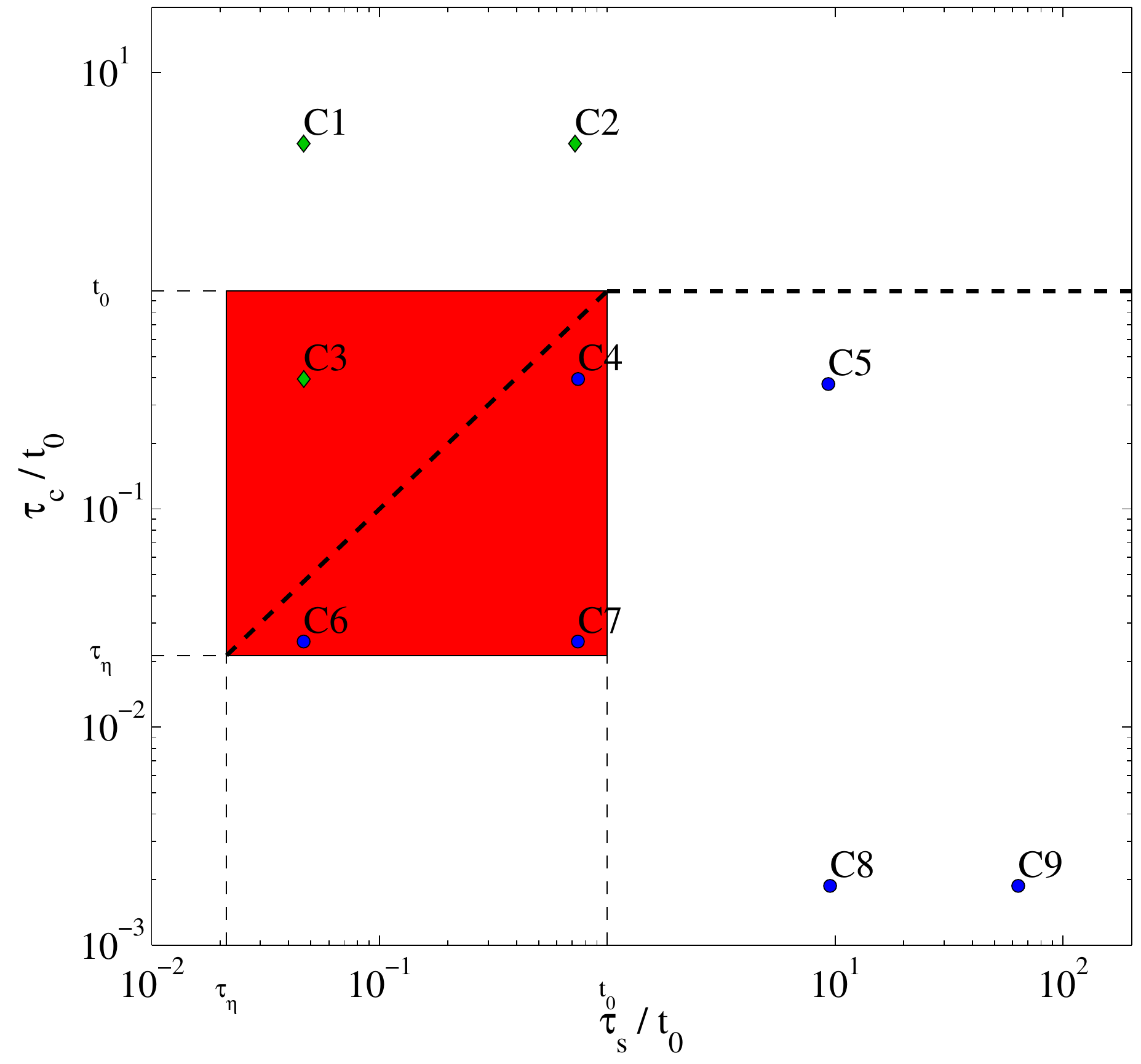} 
  }
  \caption{Plot showing the parameter space explored by DNS as labeled
    points in terms of the condensation timescale $\tau_c$ and the
    supersaturation timescale $\tau_s$.  The red area spans from the
    Kolmogorov time scale $\tau_\eta$ to the large-eddy turnover time $t_0$
    to indicate the turbulent inertial range.  The thick dashed line
    separates the region above in which
    $\tau_c > \min\left(\tau_s,t_0\right)$ with green diamonds from
    the region with $\tau_c < \min\left(\tau_s,t_0\right)$ with blue
    circles.  
}
\label{fig:param_space}
\end{figure}
\begin{figure}
  \centerline{ \includegraphics[width=0.49\textwidth]{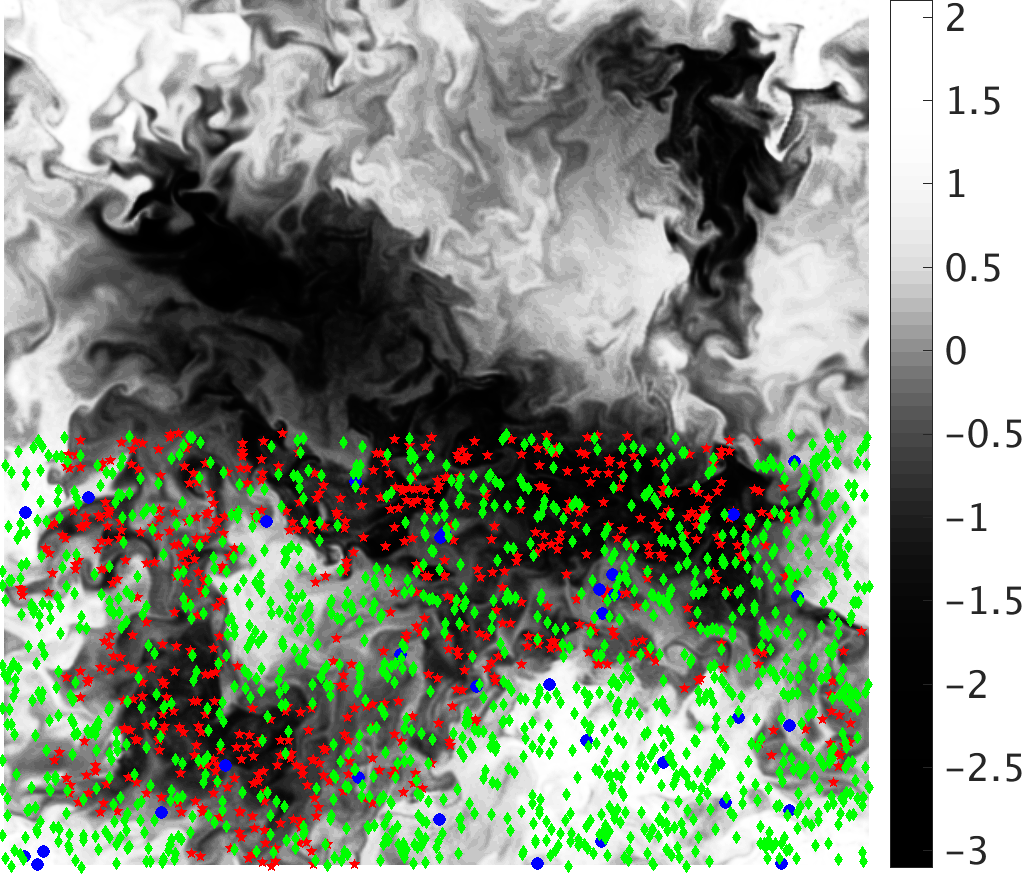}
    \includegraphics[width=0.49\textwidth]{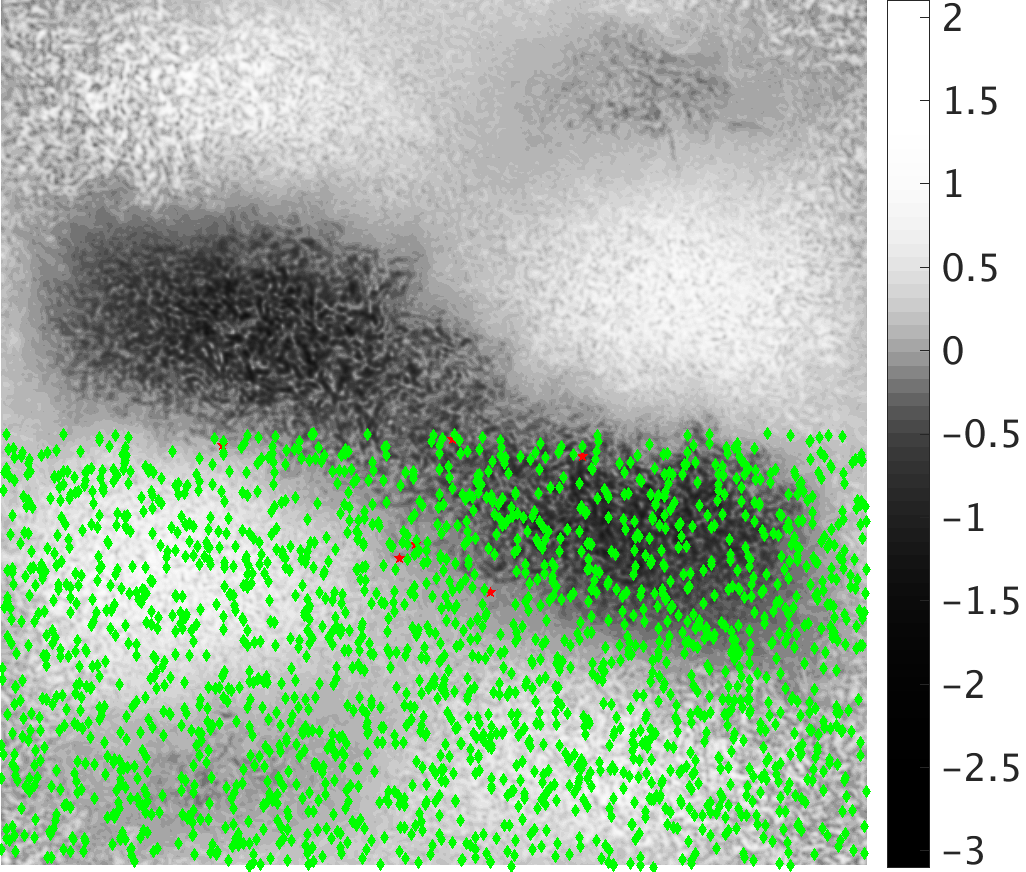}}
  \caption{ Slice of thickness $0.4~\eta$ through the 3D
    supersaturation field at $t = 3.1~t_0$; Left: Case $C4$; Right:
    Case $C3$.  The value of $s/s_{\rm rms}$ are given as grayscale
    contour plot.  Additionally the droplets are superpositioned in
    the lower half.  Their size is increased for better visibility,
    their color indicates their real size: Red stars are for
    $r^2 = 0$, green diamonds for the intermediate sizes
    ($0 < r^2 < 2\langle r^2\rangle$), and blue circles for
    $r^2 > 2\langle r^2\rangle$.  }
\label{fig:two_snapshots}
\end{figure}

\begin{table}
\begin{center}
    \begin{tabular}{r|rrrrrrrrr}
$~$ & C1 & C2 & C3 & C4 & C5 & C6 & C7 & C8 & C9 \\ \hline
$\tau_c/t_0$ & $4.737$     & $4.737$    & $0.3948$   & 0.3948 & 0.3740 & 0.02467 & 0.02467 & 0.001870 &  0.001870  \\ \hline    
$\tau_s/t_0$ & $0.04649$ & $0.7213$ & $0.04654$ & 0.7436  & 9.3253   & 0.04648 & 0.7427   & 9.491 & 63.509    
    \end{tabular}
  \end{center}
  \caption{\label{table2}
  The parameter space explored by the DNS cases C1 to C9 in terms of the condensation and
supersaturation time scales $\tau_c$ and $\tau_s$.}
\end{table}

The parameter space explored in terms of the condensation and
supersaturation time scales $\tau_c$ and $\tau_s$ is visualised in
Fig.~\ref{fig:param_space} and additionally summarized in Table~\ref{table2}.  Different combinations are chosen in or close to the turbulent inertial range in order to detect possible
interferences or resonances between the various physical processes.

In \S\ref{subsec:HandWavyingArguments} we anticipated two
  different asymptotic regimes.  For a first impression of these two
  regimes we select the cases $C3$ and $C4$ for visualisation.  They
  feature the same value of the condensation timescale $\tau_c < t_0$,
  but $C4$ has a larger value of the supersaturation timescale
  $\tau_s < t_0$, such that we expect to obtain for these instances
  the two different asymptotic regimes. Two-dimensional cuts of the
two supersaturation fields are shown in Fig.~\ref{fig:two_snapshots}.
The left-hand panel shows the case $C4$ where $\tau_c<\tau_s<t_0$. The
supersaturation field clearly looks like a turbulent passive scalar
as described in \S\ref{subsec:GoverningEqs} \citep[see][for
  a review]{warhaft2000passive}.  Hence, the influence of the
droplets is practically unnoticeable, except globally as the spatial
average of $s$ is negative. In the right-hand panel of
  Fig.~\ref{fig:two_snapshots} the case $C3$, where
$\tau_s<\tau_c<t_0$, is represented and displays a completely
different behaviour.  The local fluctuations of $s$ are strongly
reduced at both positive and negative values.  Large-scale structures
still exist, but intermediate-scale variations are smoothed out by the
droplet-vapour exchange.  At fine scales the influence of the
individual droplets can be recognized.  Their positions are
superimposed on the lower half of Fig.~\ref{fig:two_snapshots}.  There
are very few completely evaporated droplets, but no droplet that has
doubled its surface area after $3.1~t_0$. For comparison, in the case
$C4$ (left-hand panel of Fig.~\ref{fig:two_snapshots}), there are
large droplets and a huge portion of evaporated droplets can be found
in regions with a negative supersaturation.

\section{Stochastic Model}
\label{sec:model}
To understand further the different regimes discussed in
\S\ref{subsec:HandWavyingArguments}, we develop here a stochastic
model which is giving a simple approximation for the global
system~(\ref{eq:SurfaceGrowth})-(\ref{eq:SupersaturationField}).  This
model is supposed to give predictions for the global evolution of the
size distribution, rather than a good approximation for the time
evolution of single droplets. In that sense, this model should be
intended as a PDF model, rather than a pure Lagrangian approximation.
It is expected to reproduce well the one-point one-time joint
statistics of $r$ and $s$ but, in principle, cannot be used for
multiple-point or multiple-time statistics.

\subsection{Derivation}
\label{subsec:ModelDerivation}

We are interested in the
behaviour of the joint distribution of $r^2$ and $s$ along droplet trajectories at 
large times. The large-scale Stokes number being
small, we thus assume that droplet inertia can be neglected. 
Following ideas borrowed from PDF modelling of turbulent mixing
\citep[see][]{p00}, we assume that along droplet trajectories, the
variations of supersaturation due to diffusion and forcing can be
approximated at large times by a Langevin (Ornstein--Uhlenbeck) process.
The correlation time should then be the integral time scale of the 
Lagrangian scalar autocorrelation $t_s$ and the drift the 
(time-dependent) Eulerian average of the supersaturation field $\langle s\rangle_E$.
When taking
into account the variation of supersaturation due to condensation or
evaporation of the droplet from Eq.~(\ref{eq:SupersaturationField}),
this leads to model the evolution of $s$ along droplets as
\begin{equation}
  \frac{\mathrm{d}s}{\mathrm{d}t} = -\frac{1}{t_s} (s-\langle
  s\rangle_E) - \frac{1}{\tau_s}\frac{r}{\bar{r}}\,s +
  \sqrt{\frac{2 s_{\rm rms}^2}{t_s}}\,\xi(t),
\end{equation}
where $\xi$ is the standard white noise. Hence, note that this evolution equation is a stochastic differential equation, which has to be taken into account when, e.g., computing the moments of $s$.
The evolution equation is of course supplemented by
Eq.~(\ref{eq:SurfaceGrowth}) to account for the time variations of the
droplet radius $r$. This modelling involves the Eulerian average
$\langle s\rangle_E$ of the supersaturation field which is not
directly given by a Lagrangian approach but can be inferred from the
global mass conservation of liquid and vapour. We indeed have from
(\ref{eq:conser_total_mass}) combined with (\ref{eq:SupersaturationTimeScale}) and (\ref{eq:CondensationTimeScale})
\begin{equation}
w = \langle s\rangle_E + 1 +\frac{2}{3} \frac{\tau_c}{\tau_s} s_{\rm rms} = \mathrm{const},\quad\mbox{so that
} \langle s\rangle_E = \left(w-1\right) - \frac{2}{3}
\frac{\tau_c}{\tau_s}  s_{\rm rms} \propto - \langle r^3\rangle.
\end{equation}
The Eulerian averaged supersaturation is thus a global quantity that
depends on the full distribution of $r$. In that sense the system is
not closed at the level of a single droplet and the associated
Fokker--Planck equation is an integro-differential equation.  Because
of this, our model differs from that proposed by
\citet{paoli2009turbulent}. They have indeed used the Lagrangian mean
of $s$ instead of the Eulerian one.  
Additionally, we want to emphasize that the supersaturation along trajectories is 
correlated over times corresponding to the Lagrangian integral timescale of the scalar $t_s$. 
Table~\ref{table1} shows in agreement with \citet{yeung2001lagrangian} that $t_s$ 
generally differs from the large eddy turnover time $t_0$ used by \citet{paoli2009turbulent} and
the Lagrangian integral timescale of the velocity $t_u$ used by \citet{sardina2015continuous}.

To single out the relevant parameters, we next rescale time $t$ by
$t_s$, the supersaturation $s$ by $s_{\rm rms}$, and the droplet
radius $r$ by
$\bar{r}\sqrt{t_s/\tau_c} = \sqrt{2a_3 s_{\rm rms}t_s}$. 
The model then reads

\begin{subequations}\label{eq:Model}
\begin{align}
  \frac{dR^2}{dT} &= \left\{ \begin{array}{ll} S &  R^2 \geq  0 \\ 0 & R^2 = 0 \; \& \; S < 0 \end{array}  \right.  \\
  \frac{dS}{dT} &= \left( \langle W\rangle - \frac{2}{3}A\langle R^3\rangle \right) - S + \sqrt{ 2 } \xi - A R S. \label{eq:ModelS}
\end{align}
\end{subequations}
The evolution depends upon two parameters: The global mass
\begin{equation}
  \label{eq:ModelW}
  \langle W\rangle = \frac{w -1}{s_{\rm rms}} =
  \langle S\rangle_E + \frac{2}{3} A \langle R^3\rangle,
\end{equation}
and the constant $A$, which determines how strongly the evolution of $R$ couples back to the evolution of $S$. It is given by the two timescales $\tau_c$
and $\tau_s$, namely
\begin{equation} \label{eq:ModelA} A =
  \frac{t_s}{\tau_s}\sqrt{\frac{t_s}{\tau_c}} = \sqrt{32 s_{\rm rms}} \pi
  \rho_d A_2 A_3^{3/2} n_d t_s^{3/2}.
\end{equation}

\begin{figure}
  \centerline{ \includegraphics[height=0.49\textwidth]{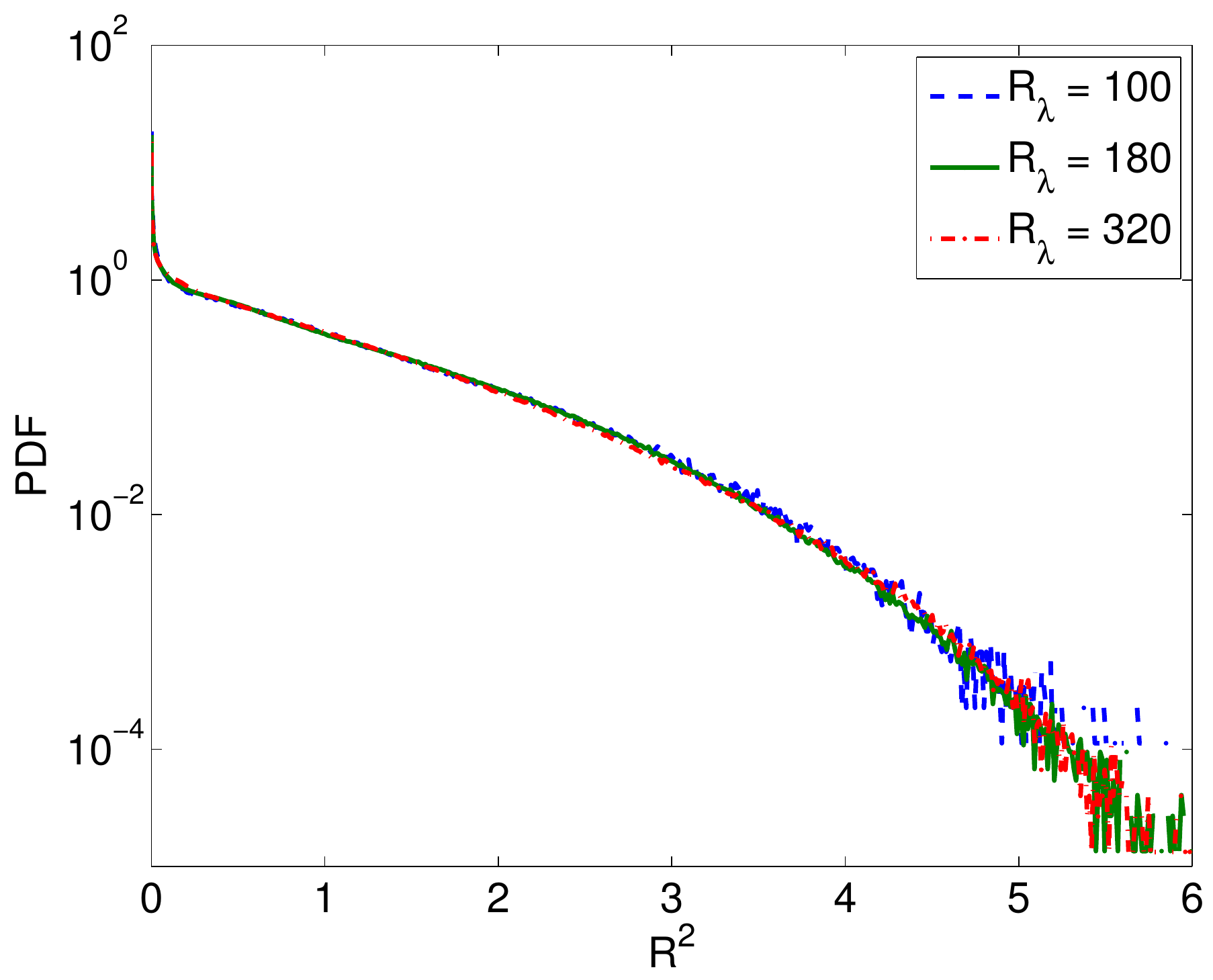}}
  \caption{Comparison of three DNS runs with increasing resolutions and therewith Taylor scale-based Reynolds numbers but same parameters $\tau_c/t_s$ and $\tau_s/t_s$, namely that of case $C8$, by the means of the PDF of $R^2$ at $t = 6.2~t_s$: Blue $256^3$ at $R_\lambda=100$; Green $512^3$ at $R_\lambda=180$; Red $1024^3$ at $R_\lambda=320$.}
\label{fig:indep_Re}
\end{figure}
To provide evidence that $t_s$ is the correct correlation time,
we conduct DNS at various resolutions, \textit{i.e.}\/ varying the Reynolds numbers, 
while at the same time we keep the dimensionless parameters $\tau_c/t_s$ and
$\tau_s/t_s$ constant.  
As depicted in Table~\ref{table1}, these parameters would differ by $20$ or $25$ percent in this Reynolds number range if $t_0$ or $t_u$ had been used instead.
Figure~\ref{fig:indep_Re} shows the
probability density function (PDF) of the normalized droplet surface area $R^2$
for the three different Reynolds numbers at a specific time T.
The collapse of the distribution confirms the role of large-scale mixing and emphasizes the choice of $t_s$ as a reference timescale. 
As $\eta$ decreases with the Reynolds number at fixed $l_0$, we also decreased $r$ accordingly to stay in the Stokes drag limit (see \S~\ref{subsec:GoverningEqs}). We kept the particle volume loading $n_d$ constant, hence, the number of particles increases with Reynolds number. That is why the $R^2$ tail contains less statistical noise at higher Reynolds number. Still, by normalizing with the Reynolds number dependent values of $t_s$ and $s_{\rm rms}$ the PDF of $R^2\left(T\right)$ is collapsing, i.e. if there is any Reynolds number dependence at all it is only very weak.

\begin{figure}
  \centerline{ \includegraphics[width=0.65\textwidth]{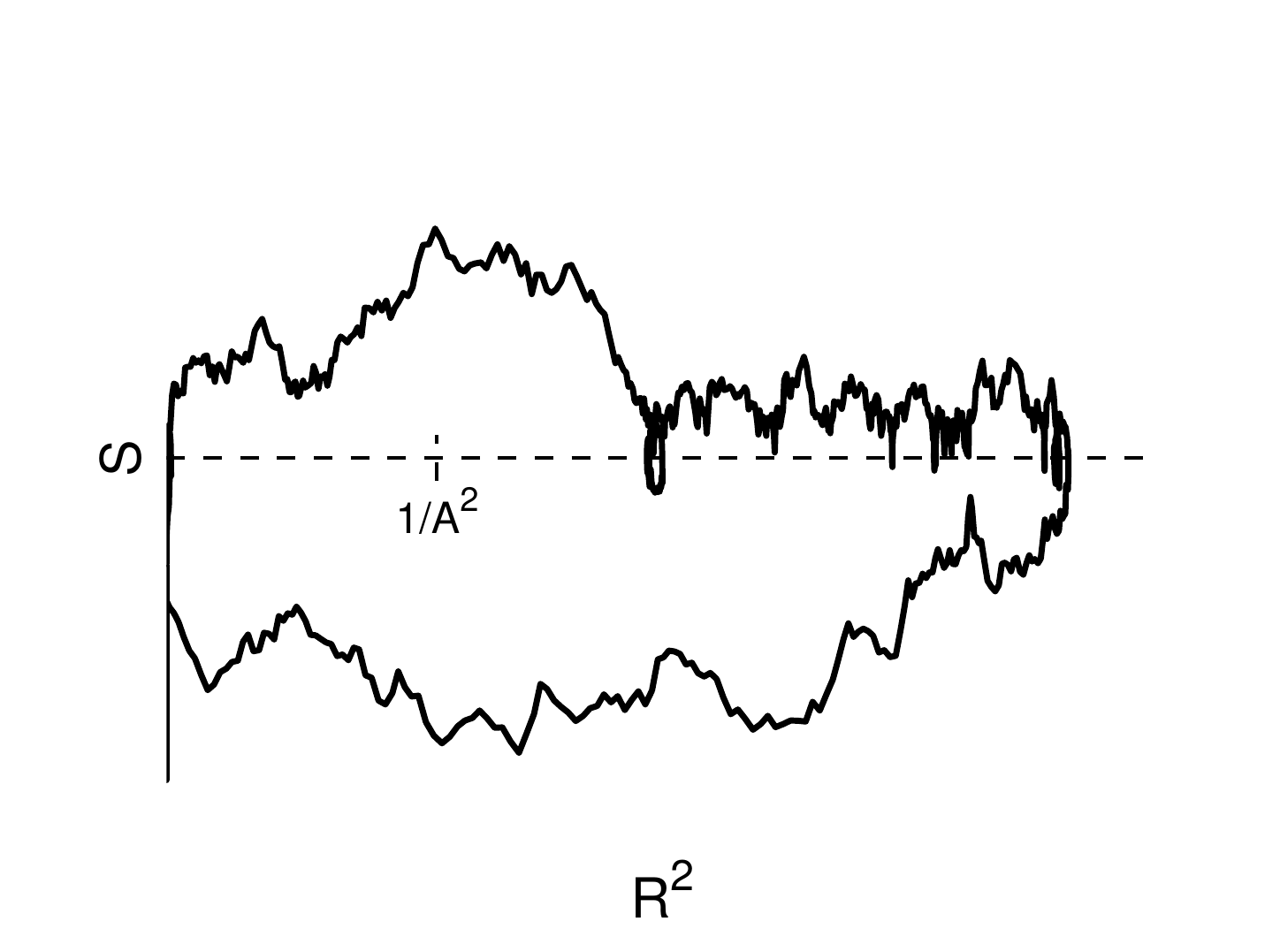}}
  \caption{Exemplary dynamics in the $R^2-S$ phase space.
}
\label{fig:PhaseSpaceDynamics}
\end{figure}

The nonlinear integro-differential stochastic
system~(\ref{eq:Model}) is much simpler than the original governing
system of partial-differential equations, but still too complex to be
solved analytically.  As done by \citet{paoli2009turbulent} the
evolution of the moments can be written down from the model
equations~\ref{eq:Model}:

\begin{subequations}\label{eq:ModelGeneralMoments}
\begin{align}
\frac{d \langle  R^{x+2} \rangle}{dT} &= \frac{x+2}{2} \langle  S R^{x}  \rangle \label{eq:ModelGeneralMomentsR} \\
\frac{d \langle  S \rangle}{dT} &= \langle S \rangle_E - \langle \left(1+AR\right)  S \rangle  \\
\frac{d \langle  S^2 \rangle}{dT} &= 2 \langle  S \rangle_E \langle  S \rangle - 2 \langle \left(1+AR\right) S^2 \rangle + 2 \\
\frac{d \langle  S R^2 \rangle}{dT} &= \langle S \rangle_E \langle  R^2 \rangle + \langle S^2 \rangle - \langle \left(1+AR\right)  S R^2 \rangle \quad .
\end{align}
\end{subequations}
This is an unclosed hierarchy of equations and can only be solved if additional closure assumptions are made.
However, it is possible to understand qualitatively the system dynamics.
A typical trajectory in the phase
space is shown in Fig.~\ref{fig:PhaseSpaceDynamics}.  
Due to the noise term the supersaturation tends to do Brownian
excursions.  For negative values of $S$ the droplet shrinks while for
positive values the droplet grows.  The drift term prevents runaway
excursions to $s=\pm \infty$.  Note that the non-linear coupling term
$-ARS$ is dominant for large droplets $R^2 \gg 1/A^2$.  Thereto the
larger the droplet is, the faster a positive $S$ is reduced to zero,
so that large droplets can growth only on very short timescales.
Conversely, an evaporating droplet is less and less effective in
pushing its negative supersaturation back to zero and thus has longer
and longer time to shrink.  Due to this bias it seems reasonable to
assume that a droplet cannot grow infinitely large.  In the opposite
limit, when a droplet completely evaporates (reaches $R=0$), the
dynamics of $S$ decouples from $R$ and the droplet will start growing
only once $S$ becomes positive, reinitializing the full process again.
There is thus a loss of memory of the previous growth history.
Bounded excursions, together with the recurrent memory losses suggest
that, at long time, the system should reach a statistical stationary
state independent of its initial condition.

\subsection{Comparison with DNS}\label{subsec:ModelDNSComparison}

\begin{figure}
  \centerline{ \includegraphics[width=0.95\textwidth]{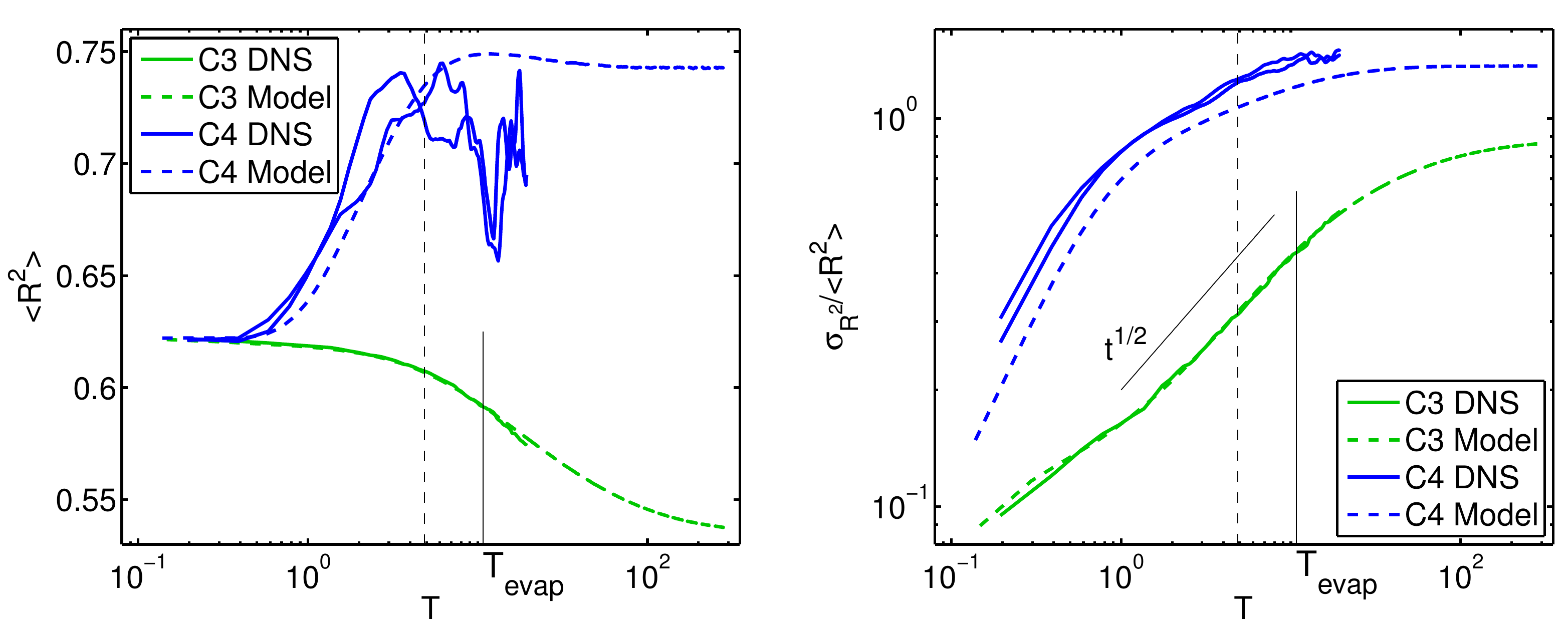}}
  \caption{Comparison of the DNS (solid lines) with the results of the
    stochastic model (dashed lines) for the cases $C3$ (green) and
    $C4$ (blue).  Note that for case $C4$ two independent DNS runs are
    shown.  Left: Time evolution of the mean droplet surface area
    $\langle R^2\rangle$; Right: Time evolution of the relative
    standard deviation of the droplet surface area
    $\sigma_R^2/\langle R^2\rangle$.  The vertical black solid line
    marks $t_{evap}$ for $C3$ at which $\sigma_R^2/\langle R^2\rangle$
    is predicted to start deviating from the $t^{1/2}$ behavior (see
    \S\ref{subsec:HandWavyingArguments} for the preliminary
    heuristic arguments and sec.~\ref{subsec:ShortTimeBehavior} for a detailed
    derivation of eq.~(\ref{eq:Tevap})).  The vertical black dashed line marks the time of the
    snapshots shown in Fig.~\ref{fig:two_snapshots}.  }
\label{fig:compare_convergence}
\end{figure}

We perform Monte-Carlo (MC) simulations of the system (\ref{eq:Model})
for the same parameters as in the DNS cases $C3$ and $C4$ shown in
Fig.~\ref{fig:two_snapshots}.  Thereto the parameters
$A$~(\ref{eq:ModelA}) and $\langle W \rangle$~(\ref{eq:ModelW}) are
calculated from the DNS initial conditions. Especially $t_s$ and
$s_{\rm rms}$ are measured from the initial supersaturation field as
the afterwards released droplets change its properties.  The
parameters $A$ and $\langle W \rangle$ both depend on $s_{\rm rms}$,
hence, the model is sensitive to the value of $s_{\rm rms}$.  The MC
simulations allow us to simulate the system for much longer times than
for the DNS.

The evolution of $\langle R^2\rangle$ is shown in the left panel of
Fig.~\ref{fig:compare_convergence}.  In the case $C3$ the mean value
of $R^2$ for both the DNS and the model, first remains relatively
constant and then even decreases.  In contrast, for the case $C4$, the
average of $R^2$ increases substantially before converging to a
constant value.  In this case, the model lags a little bit behind the
DNS. Additionally, the final value seems to be a few percent higher,
although this is not unambiguous as the DNS value strongly fluctuates.
That is why we have represented two different realizations of the full
DNS C4 in Fig.~\ref{fig:compare_convergence}. Given the large
amplitude of the variations between these two runs and of the
fluctuations as a function of time, the deviations from the model
could be explainable by a lack of ensemble averaging.  While the
number of droplets is the same for the DNS and the MC simulation, the
supersaturation in DNS is correlated on the large scales, whereas the
random increments are independent in the model.  Hence, the
number of independent realizations is higher in the model.  Ensemble
averaging of independent DNS realizations would be needed but is
computationally too expensive.
 
The right panel of Fig.~\ref{fig:compare_convergence} represents the
coefficient of variation of the droplet surface area, namely the
relative standard deviation $\sigma_R^2/\langle R^2\rangle$.  In the
case $C4$ it increases first rapidly and then converges, although more
slowly than the mean value, to a value close to one.  The model values
are slightly lower than the DNS.  Nevertheless, for both the model and
the DNS, the coefficient of variation is order one, indicating a very
broad distribution close to the one given for instance by an
exponential distribution. For the case $C3$, the relative standard
deviation starts much lower and for a decade of time, the predicted
$t^{1/2}$ power-law can be seen in the logarithmic plot. Hence,
while the case $C4$ is already in the steady state in
Fig.~\ref{fig:two_snapshots}, $C3$ is still in the Brownian motion
regime, leading to the very different pictures.  However, also in the
case $C3$, the model eventually converges to a constant value,
\textit{i.e.}\/ a steady state is reached at very large times.

To conclude we want to emphasize two points: First, the MC simulations
confirm the above reasoning that, at long times, the system converges
to a steady state.  Hence, the heuristic arguments of
\S\ref{subsec:HandWavyingArguments} actually describe two regimes of
the transient system behaviour.  Second, the model predictions are
representative of the actual dynamics.  The simple stochastic model is
indeed able to reproduce some results of the DNS surprisingly well.
This is unexpected as the turbulence is only modeled trivially and no
intermittency or inertia effects are taken into account.  The
short-time behaviour is perfectly reproduced, while there are
deviations for the steady state.  However, these deviations are
systematic for all DNS runs reported in Fig.~\ref{fig:param_space}
that we have conducted. We will come back to this issue in the next
section.

\section{The Steady State}
\label{sec:SteadyState}

Here we characterise the steady state by analytical predictions based on the
model, compare these with the numerical results and show the
dependence of the steady state on the two model parameters.

\subsection{Model-based Analytical Predictions}\label{subsec:TheoResults}
We can find from the equations of the
moments~(\ref{eq:ModelGeneralMoments}) that in the steady state, the
Lagrangian and Eulerian mean of the supersaturation coincide
$\langle S \rangle = \langle S \rangle_E$ as all time derivatives are
zero and thus the mixed moments $\langle S R^{x}\rangle$ all vanish.
The latter can only be fulfilled if the probability density function
of the supersaturation $s$ conditioned on any $R>0$ is symmetric
around zero.  Due to the convergence of the mean droplet mass
$\langle R^3\rangle$ and of the mean Eulerian supersaturation
$\langle S\rangle_E$, the system (\ref{eq:Model}) is no longer a
stochastic integro-differential equation as it does not involve
anymore integrals.  Hence, it can be described by its associated
stationary Fokker--Planck equation.  To reflect the piecewise
definition of the stochastic model (\ref{eq:Model}), we split the
joint probability density $p$ into an atomic contribution $m$ for
evaporated droplets and a smooth part $\tilde{p}$ for active droplets
\begin{equation}
  p\left(S,R^2,T \right)  = 
  m\left(S,T\right)\,\delta(R^2) + \tilde{p}\left(S,R^2,T\right) \,.
  \label{eq:SplitPDFs}
\end{equation}
The stationary Fokker-Planck equation for the active droplets
$\tilde{p} \left(S,R^2, T \rightarrow \infty \right)$ can then be
written as
\begin{equation}
  \frac{\partial}{\partial R^2}\left( S \tilde{p} \right) +
  \frac{\partial}{\partial S}\left( \left[\langle S\rangle_E - \left(1 +
        A R\right) S - \frac{\partial}{\partial S} \right] \tilde{p}
  \right) = 0\,.
  \label{eq:FokkerPlanck}
\end{equation}
Despite the simplifications in the steady state, finding a general
solution is not straightforward  
 because of the non-linear coupling term
$\left(1+AR\right)S$.  Nevertheless, we can predict the shape of the
tail of the distribution $p(R^2)$ of the droplet surface area.
Interpreting the surface area $R^2$ as position and the
supersaturation $S$ as velocity, equation~(\ref{eq:FokkerPlanck}) is
similar to the Fokker--Planck equation associated with Brownian
particles subject to a position-dependent drag.  In the large-drag
limit the velocity is a fast variable and can be eliminated.
Following \citet{sancho1982adiabatic}, an expansion of the non-linear
drag term yields to leading order
\begin{equation}
  \label{eq:SteadyStateRPDF}
  p(R^2) \simeq - \langle S\rangle_E
  \left(1+AR\right)\exp\left( \langle S\rangle_E R^2 + \langle
    S\rangle_E \frac{2}{3} A R^3 \right) .
\end{equation}
This solution is only valid when the term $\left(1+AR\right)$ is
dominant, \textit{i.e.}\/ for large droplets with $R^2 \gg 1/A^2$ (see
Fig.~\ref{fig:PhaseSpaceDynamics}).  One could think of finding
another solution for the opposite asymptotic $R^2 \ll 1/A^2$ and then
match the two limiting solutions similarly to the approach in boundary
layer theory.  However, due to the finite probability of having
completely evaporated droplets, see Eq.~(\ref{eq:SplitPDFs}), and the
complicated boundary conditions at $R=0$, this is a very challenging
task.  Still, the large-droplet solution (\ref{eq:SteadyStateRPDF})
has global consequences: To have an integrable distribution of large
droplets the mean Eulerian supersaturation $\langle S \rangle_E$ has to
be negative.  As seen above, $S$ conditioned on $R>0$ has to be
symmetric around zero, such that the mean value of $S$ in the presence
of droplets is zero.  Hence, only $S$ conditioned on $R=0$ contributes
to the negative $\langle S \rangle_E$.  Therefore, it follows that the
presence of completely evaporated droplets is a prerequisite for
reaching the steady state.

\subsection{Comparison with Numerical Results}
\label{subsec:SteadyStateComparison}
\begin{figure}
  \centerline{ \includegraphics[width=0.95\textwidth]{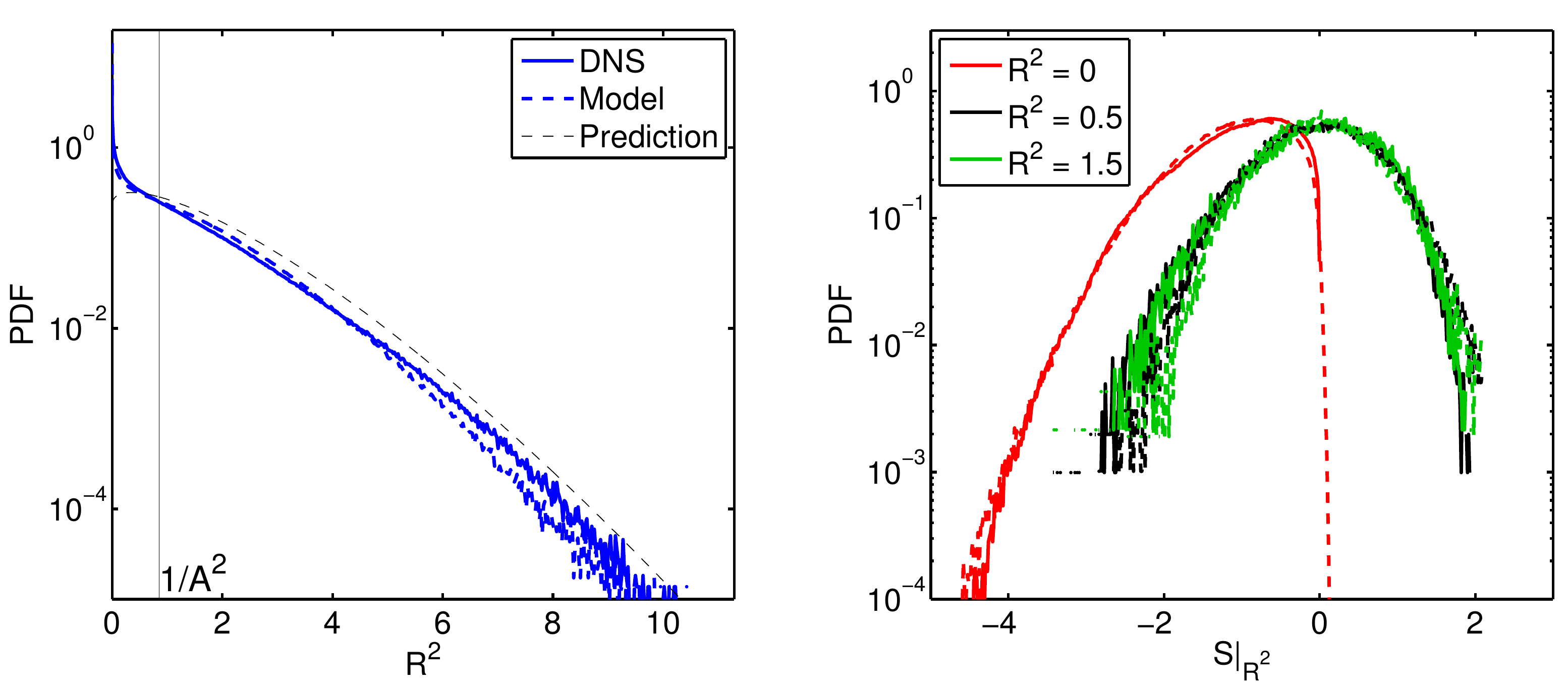}}
  \centerline{ \includegraphics[width=0.95\textwidth]{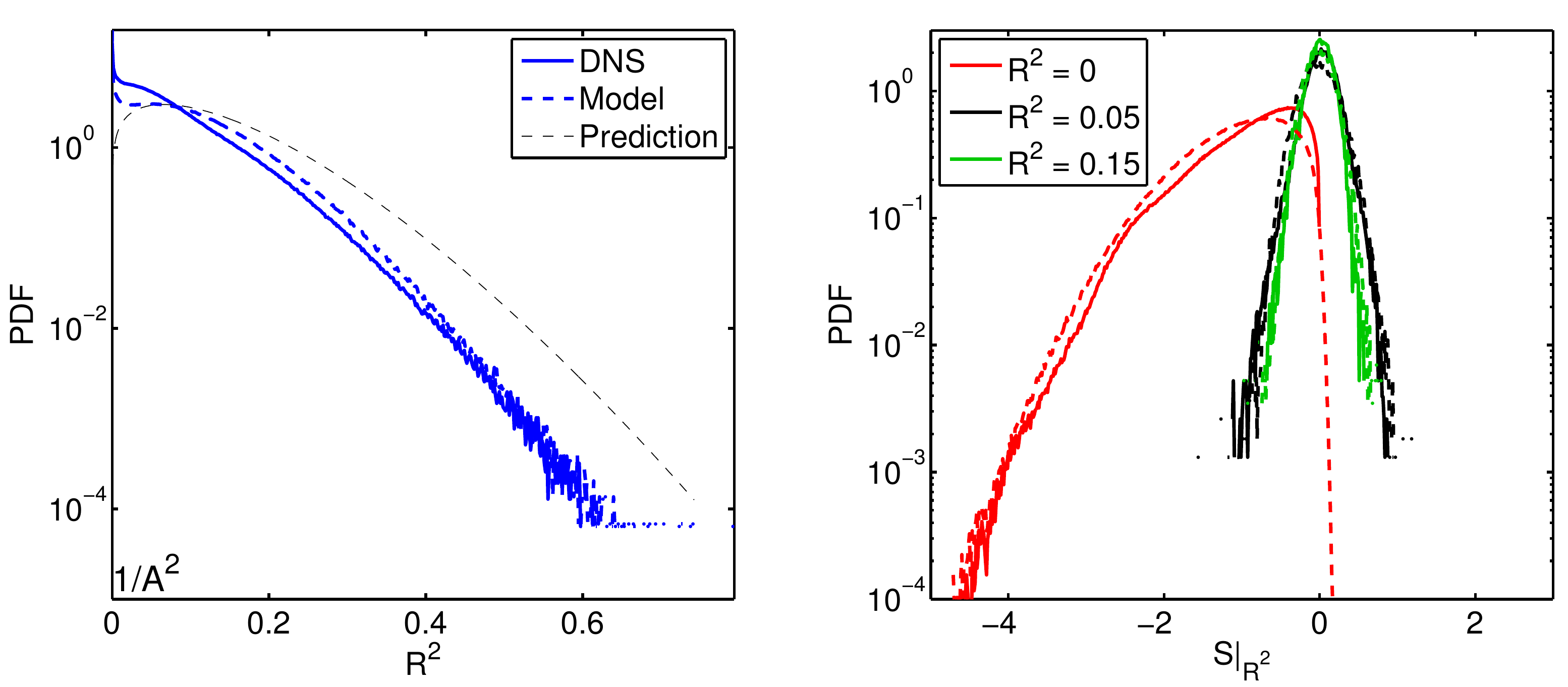}}
  \caption{Left side: Probability density function of $R^2$ from the
    DNS (solid line) and the model (dashed line).  The prediction from
    Eq.~(\ref{eq:SteadyStateRPDF}) is shown as thin dashed line.  The
    value $1/A^2$ is indicated as a thin solid black line.  Right
    side: Probability density function of $S$ conditioned on three
    values of $R^2$.  Again, the DNS results are drawn as solid lines
    while the results from the model are shown as dashed lines.  The
    PDFs are computed for the case $C4$ (top row) and the case $C6$
    (bottom row).}
\label{fig:pdf_R_s}
\end{figure}

In the following we show that the theoretical predictions for the PDFs
of $R^2$ and $S$ not only apply to the model but also to the DNS.  In
Fig.~\ref{fig:pdf_R_s} the predictions, the MC simulations, and the
DNS are contrasted by the means of the PDF of $R^2$ and the
conditioned PDFs of $S$.  For the case $C4$, which is presented in the upper half of Fig.~\ref{fig:pdf_R_s}, we know from
Fig.~\ref{fig:compare_convergence} that the model predicts a slightly
higher $\langle R^2\rangle$ than the DNS. The PDF of $R^2$ obtained
from the model shows a higher probability for intermediate-size
droplets, which leads to this higher mean value.  For $R^2 \gg 1/A^2$
the MC simulation fits the tail prediction (\ref{eq:SteadyStateRPDF}).
The tail of the PDF of the DNS is parallel to this prediction,
\textit{i.e.}\/ the shapes match, up to a multiplicative constant.
The PDF of the supersaturation $S$ is shown conditioned on the droplet
size.  For evaporated droplets ($R^2=0$) the PDF of $S$ has only
negative values.  For finite-size droplets the PDFs are centered at
zero and nearly symmetric with practically constant variances,
independent of the value of $R^2$ on which the PDF is conditioned.
The model fits well the supersaturation PDFs of the DNS though some
small deviations can be observed in the negative-$S$ tail.

The trends discussed in the case $C4$ are representative for all DNS
runs that we have conducted and are reported in
Fig.~\ref{fig:param_space}.  To corroborate this, we also show the
data corresponding to the case $C6$ in the lower half of Fig.~\ref{fig:pdf_R_s}, which features smaller $\tau_c$
and $\tau_s$, \textit{i.e.}\/ a larger $A$.  The PDF of $R^2$ obtained
with the model coincides with the prediction at large droplet sizes.
Intermediate $R$'s are less probable in the DNS but the shape of the
tail fits.  The PDFs of $s$ are well reproduced.  While the PDF of $S$
for evaporated droplets looks very similar to that for the case $C4$,
the active droplets damp the supersaturation more strongly due to the
larger coupling parameter $A$.  Because of a smaller variance of $S$
the droplets are smaller in the case $C6$ than in the case $C4$.

\subsection{Characterisation By The Model Parameters}
\label{subsec:SteadyStateCharacterization}

\begin{figure}
  \centerline{ \includegraphics[width=0.65\textwidth]{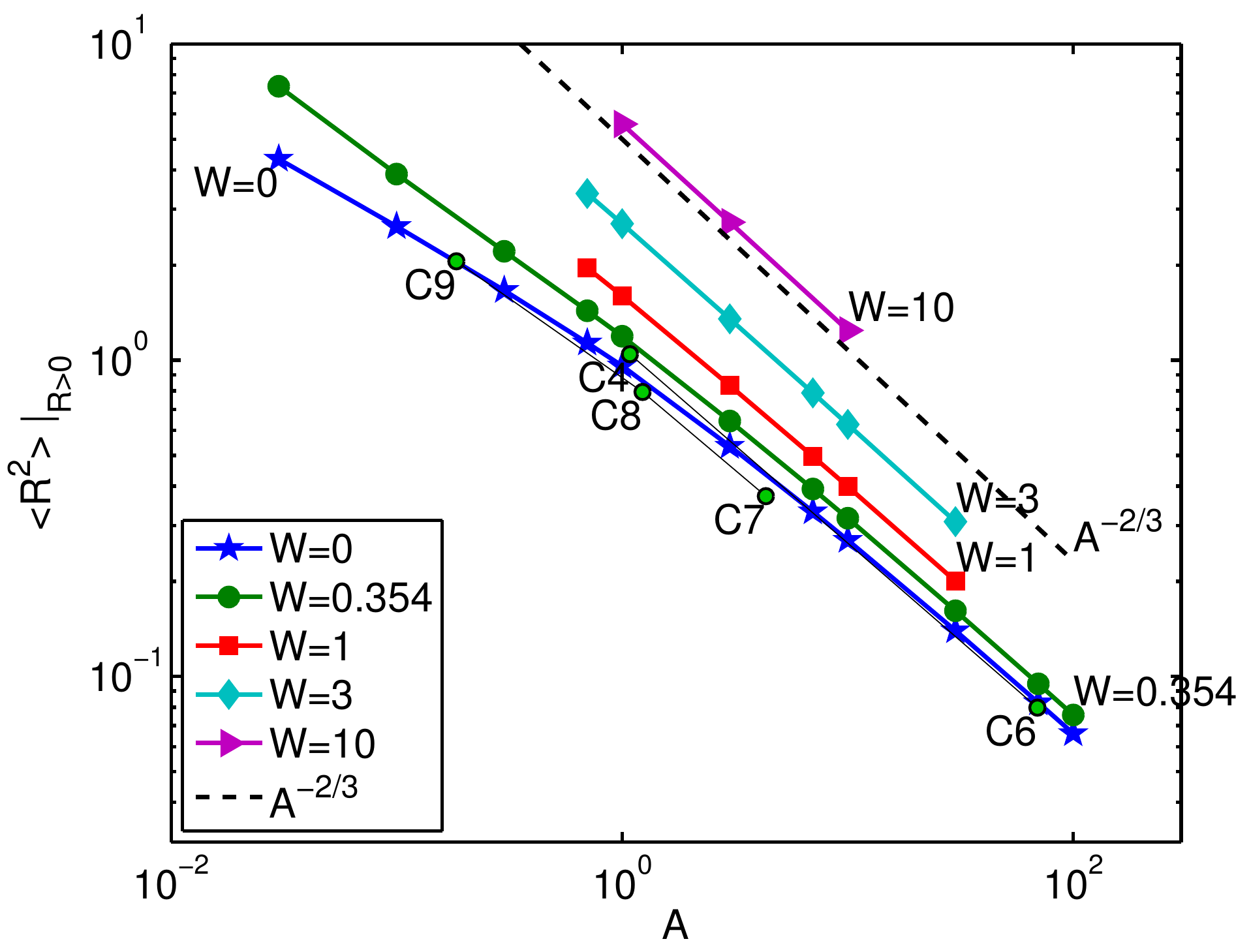}}
   \centerline{ \includegraphics[width=0.65\textwidth]{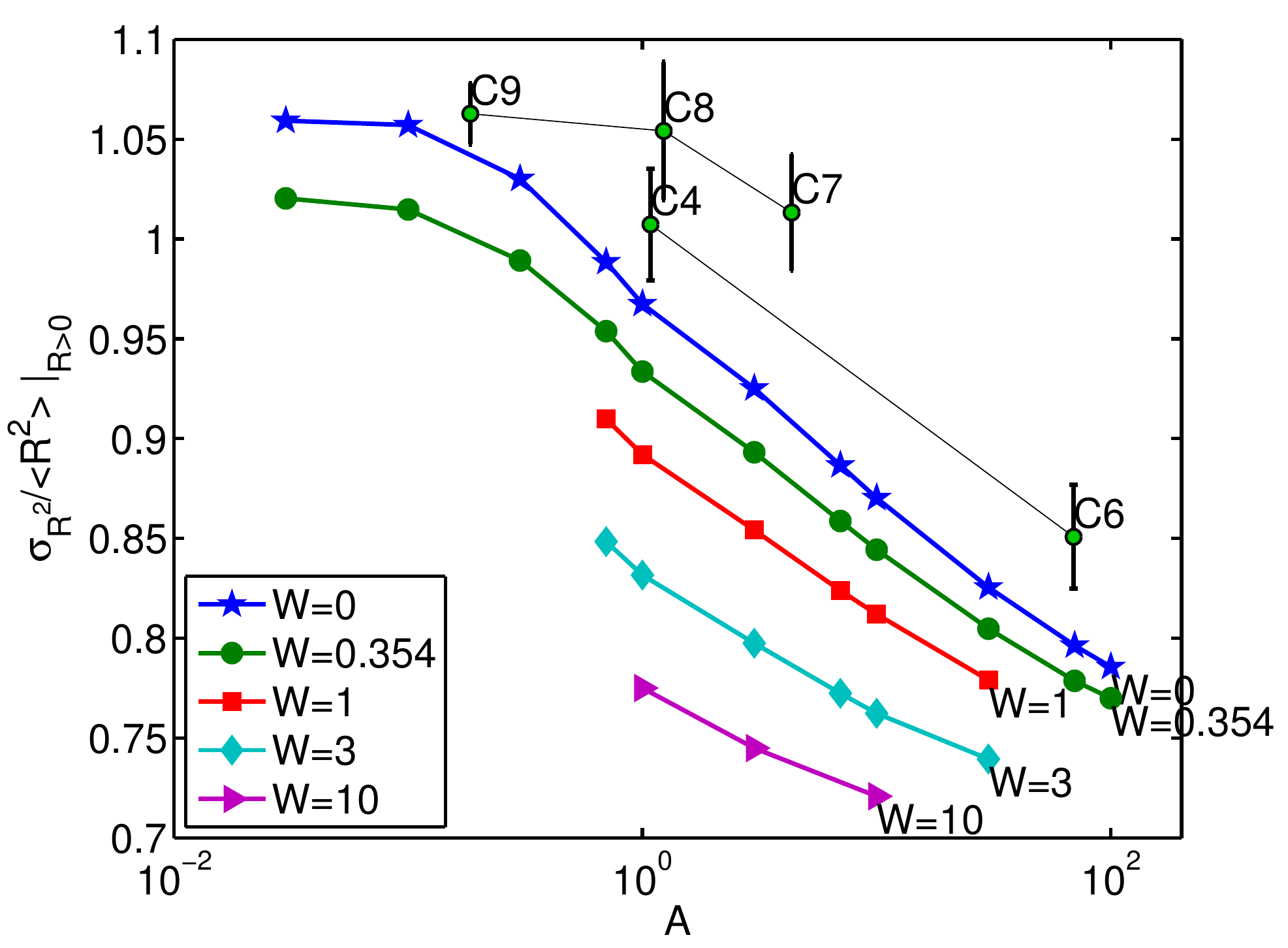}}
   \centerline{ \includegraphics[width=0.65\textwidth]{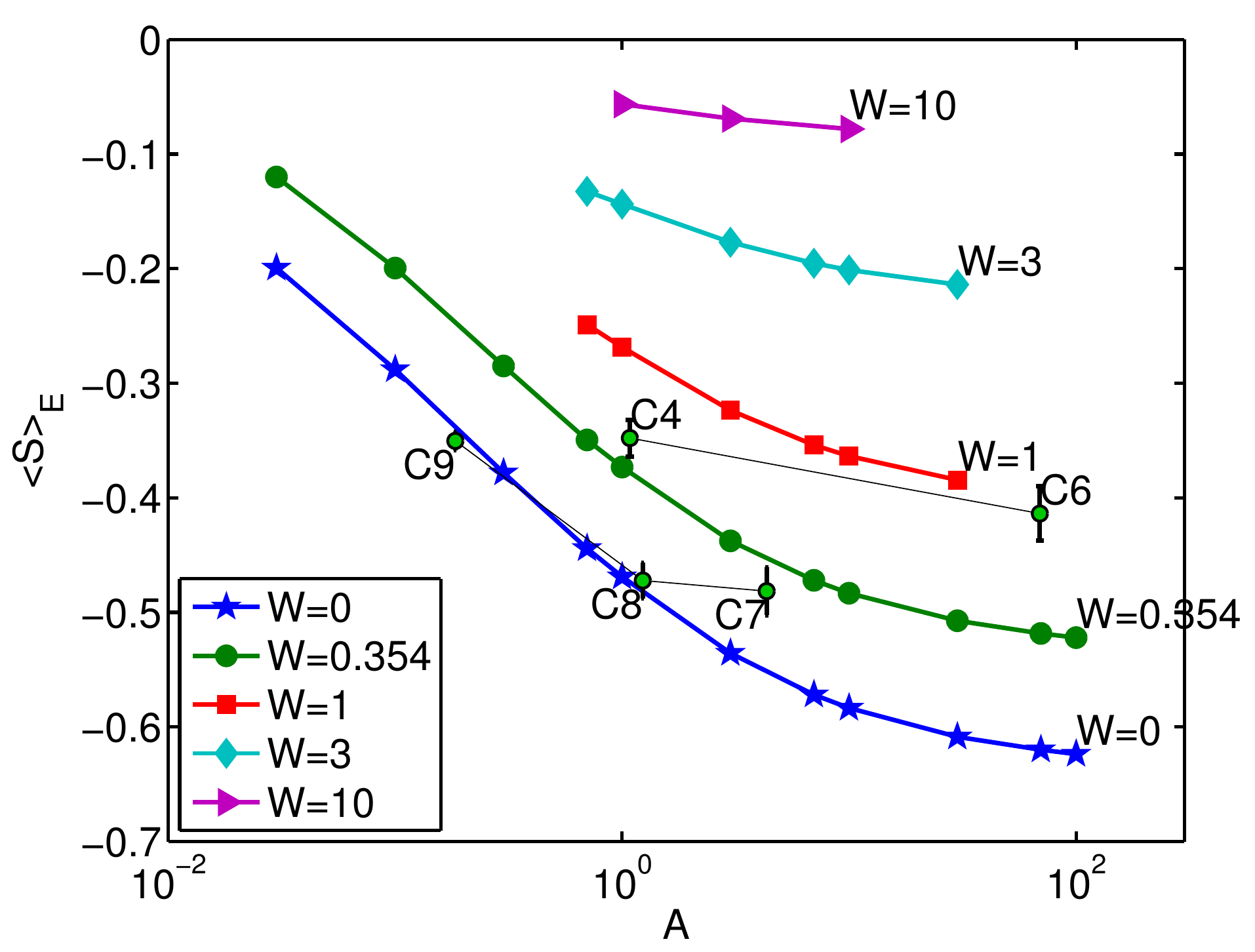}}
   \caption{Steady-state values of the mean drop surface area
     $\langle R^2\rangle |_{R > 0}$ (top), the relative standard deviation
     $\sigma_{R^2}/\langle R^2\rangle |_{R > 0}$ (middle), and the mean Eulerian
     supersaturation $\langle S\rangle_E$ (bottom) as a function of
     $A$ and with the water parameter $\langle W\rangle$ ranging from
     $0$ to $10$.  For comparison with these model results the DNS
     values of the cases $C7$, $C8$, and $C9$ with
     $\langle W\rangle \sim 0$ and $C4$ and $C6$ with
     $\langle W\rangle \sim 0.354$ are superimposed.}
\label{fig:steady_sate_fn_a}
\end{figure}
In the steady state the system (\ref{eq:Model}) depends only on the
two model parameters, namely the coupling parameter $A$ defined in
(\ref{eq:ModelA}) and the total mass $\langle W\rangle$ associated with
the conservation of liquid and vapour (\ref{eq:ModelW}).  In
Fig.~\ref{fig:steady_sate_fn_a} the steady state is characterised by
the mean drop surface area, its relative standard deviation, and the
mean Eulerian supersaturation which are represented as a function of
$A$ for various values of $\langle W\rangle$.

The mean drop surface area $\langle R^2\rangle |_{R > 0}$, displayed at the top of 
Fig.~\ref{fig:steady_sate_fn_a}, shows approximately a
power-law behaviour $\propto A^{-2/3}$.  This can be explained by the
conservation of mass (\ref{eq:ModelW}).  For large $A$ the third
moment of $R$ corresponding to the mean drop mass has to decrease
correspondingly.  This argument also explains why $\langle R^2\rangle |_{R > 0}$
increases with $\langle W\rangle$.  We would like to emphasize the
fact that mean-field arguments predict no drop growth for an initially
vanishing average supersaturation.  In contrast, due to the
supersaturation fluctuations $\langle R^2\rangle |_{R > 0}$ always converges to
a positive, non-zero value even for zero total mass in the system.

Following the same scaling arguments $\langle R^4\rangle$ is
proportional to $A^{-4/3}$, such that, to leading order,
$\sigma_{R^2}/\langle R^2\rangle |_{R > 0}$ is expected to be constant.
However, one finds in the middle of 
Fig.~\ref{fig:steady_sate_fn_a} that the size broadening slightly decreases with
$A$.  This can be explained by the tail prediction, since
$\langle R^4\rangle$ depends strongly on the large droplets.  For
large $A$, the droplet size distribution has a sub-exponential tail given
by (\ref{eq:SteadyStateRPDF}).  Consistently, the relative standard
deviation features values smaller than $1$, which decrease slowly with
$A$.  For $A$ going to zero, an exponential distribution would be
expected from Eq.~(\ref{eq:SteadyStateRPDF}), similarly to the case of
Brownian motion with reflection.  However, the shape prediction is
only valid for $\left(1+AR\right)$ large and hence the moments become
dominated by the unknown small-drag distribution.
The mean drop size increases with $\langle W\rangle$, so that the tail
prediction is valid for a larger portion of the PDF of $R^2$ and
$\sigma_{R^2}/\langle R^2\rangle$ becomes smaller.

A negative mean Eulerian supersaturation $\langle S\rangle_E$ is a
requirement for a steady state (see
\S\ref{subsec:TheoResults}).  
Note that 
\begin{equation}
\langle S \rangle_E = \langle S \rangle = \langle S | R = 0\rangle p\left(R=0\right) + \int \langle S | R\rangle p\left(R\right) dR = \langle S | R = 0\rangle \times \frac{N_{evap}}{N}\;.
\end{equation}
Since the PDF of $S$
conditioned on $R=0$ seems to be independent of both $A$ and
$\langle W\rangle$ (see Fig.~\ref{fig:pdf_R_s}), the value of
$\langle S\rangle_E$ is directly proportional to the number of
evaporated drops.
For a small coupling parameter $A$, the tendency
for droplets to be pushed towards $R^2=0$ is low, so that most of them
are active (see lower part of Fig.~\ref{fig:steady_sate_fn_a}). 
As $A$ increases, the steady-state fraction of evaporated
droplets increases.  This leads to more negative contributions to
$\langle S\rangle_E$.  The mean size $\langle R^2\rangle$ increases
with $\langle W\rangle$, and therefore the distance to the zero-size
boundary becomes larger.  Hence, for larger $\langle W\rangle$, the
probability for evaporated droplets is smaller and
$\langle S\rangle_E$ is larger, closer to zero.

The DNS statistics are also shown in Fig.~\ref{fig:steady_sate_fn_a}.
These values are less certain, since, as visible in
Fig.~\ref{fig:compare_convergence}, they fluctuate more strongly and
the fourth-order moment of $R$ might not be fully converged to its
steady-state value at the end of the simulation time of $20\,t_s$.  As
expected from Fig.~\ref{fig:compare_convergence}, the DNS cases show
lower mean values and higher relative standard deviations.
Nevertheless, their scalings are consistent with the model results.

\begin{figure}
  \centerline{ \includegraphics[width=0.49\textwidth]{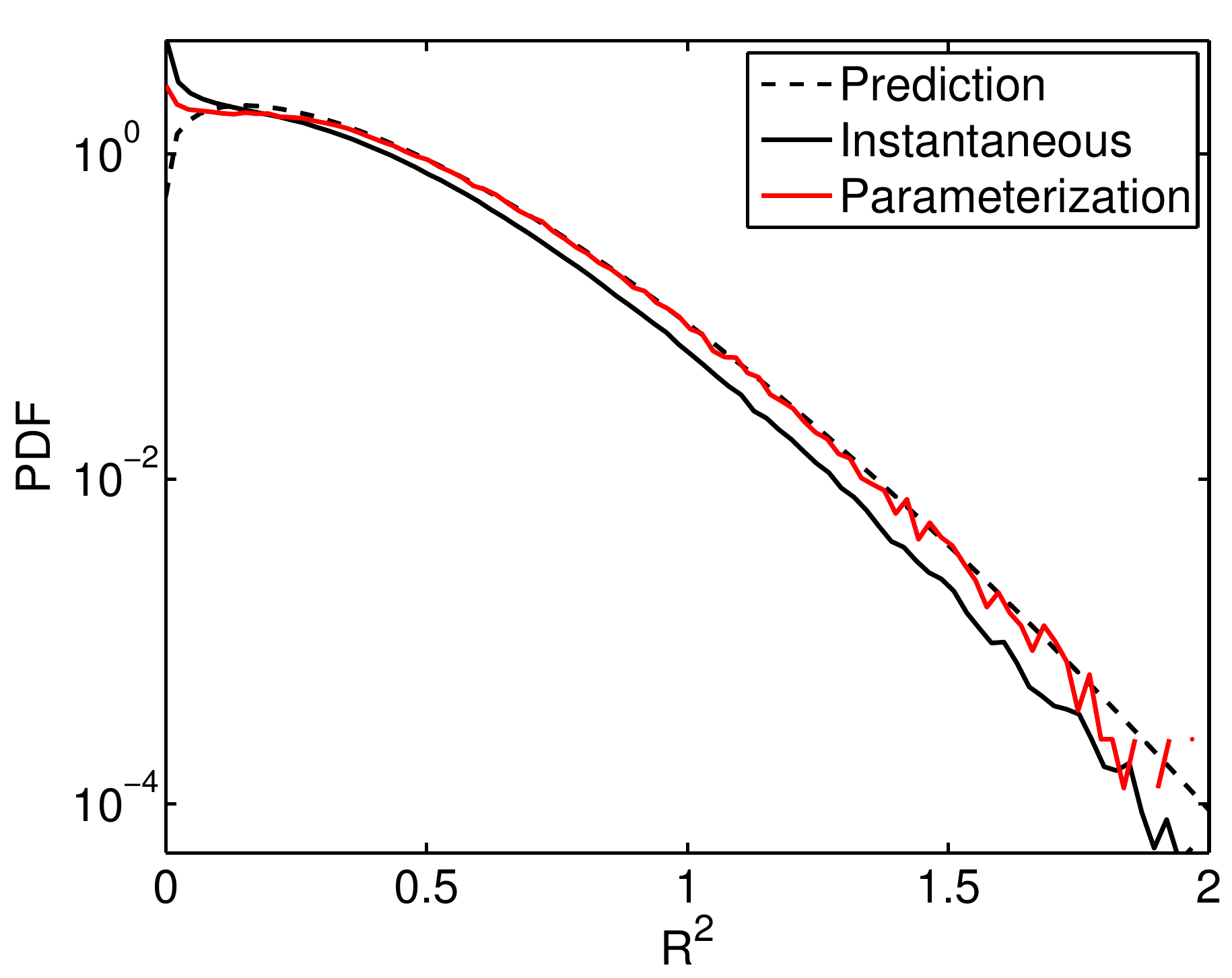}}
  \caption{Comparison of the steady-state $R^2$ PDF of two MC
    simulations with different boundary conditions at $R=0$. Black
    solid line: Instantaneous reactivation as soon as the
    supersaturation $S$ is positive, according to
    Eq.~(\ref{subsec:GoverningEqs}); Red solid line: Activation
    according to a power-law parameterisation in $S$
    \citep{seifert2006two}; Black dashed line: Prediction given by
    Eq.~(\ref{eq:SteadyStateRPDF}).  }
\label{fig:indep_BC}
\end{figure}
Finally, to show that the shape of the PDF of $R^2$ at large droplet sizes does
not depend on the treatment of those with a size close to zero, the
model is modified to activate droplets according to a parameterisation
given by the K\"ohler-Kelvin theory \citep[see,
\textit{e.g.},][]{seifert2006two}.  In this parameterisation the
activation probability behaves as a power-law on $S$.  
Therefore the
time a droplet spends at $R=0$ until it is reactivated is increased.
Hence, the number of evaporated drops in the steady state increases.
However, as it can be seen in Fig.~\ref{fig:indep_BC}, the dynamics of
large drops faraway from the boundary is practically unaffected.  This
is consistent with the findings of \citet{celani2008equivalent} (see
\S\ref{sec:intro}).

\section{Transients}
\label{sec:transients}
Depending on the physical situation, the time to reach the
steady state can be too long compared to the time of interest such
that the transient behaviour is of interest.  As anticipated in the
context of Fig.~\ref{fig:compare_convergence}, the time evolution of
the system (\ref{eq:Model}) can be differentiated into two stages, a
short-time and a long-time behaviour.  In the following these two
transients are described.

\subsection{Short-time Behaviour\label{subsec:ShortTimeBehavior}}

\begin{figure}
  \centerline{ \includegraphics[width=0.99\textwidth]{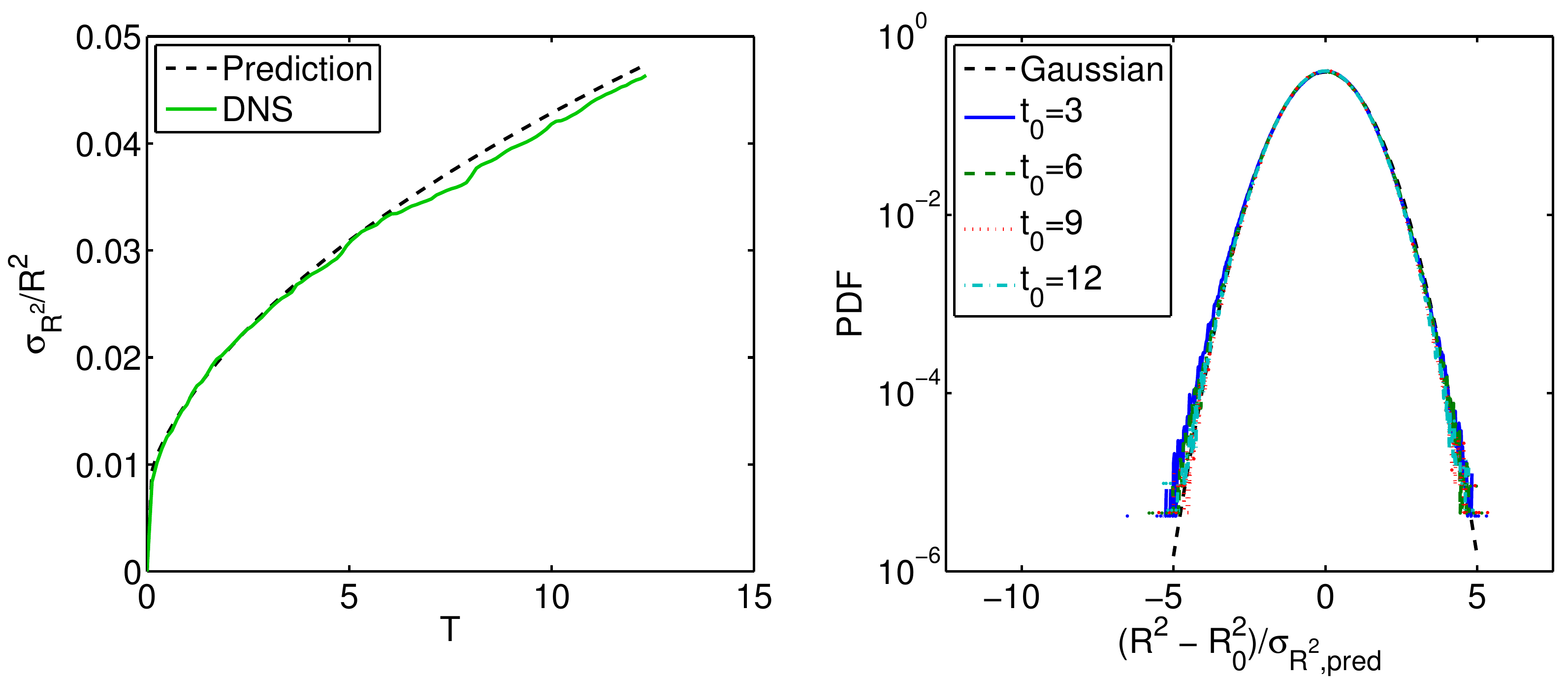}}
  \caption{Short-time behaviour of the DNS in the case $C1$.  Left:
    Relative standard deviation $\sigma_{R^2}/\langle R^2\rangle$ as a
    function of time (solid line), in comparison to the prediction of
    Eq.~(\ref{eq:ModelR2ShortTime}) (dashed line).  Right: PDF of
    $R^2$ standardized by the initial mean value $\bar{R}_0^2$ and the
    standard deviation $\sigma_{R^2,pred}$ predicted by
    Eq.~(\ref{eq:ModelR2ShortTime}) at different times (solid lines)
    and in comparison with a standard Gaussian distribution (dashed
    line).}
\label{fig:ShortTimeBehavior}
\end{figure}

At short times the evolution of the system still depends on the
initial conditions for $S$ and $R$.  We exemplarily explain the
short-time behaviour of Eqs.~(\ref{eq:Model}) by concentrating on the
DNS settings with a normally distributed supersaturation
$S \sim \mathcal{N}(0,1)$, where $\mathcal{N} \left( \mu, \sigma^2 \right)$ denotes a normal distribution with mean $\mu$ and a standard deviation $\sigma$ and a monodisperse drop spectrum
$\delta\!\left(R-R_0\right)$.  Let us assume that, at short times,
the mean droplet mass $\langle R^3\rangle$ and the coupling
  $\left(1+AR\right)$ remain approximately constant.  The second
  assumption might seem a bit crude as we are interested in the
  evolution of the individual radii $R$.  However, turbulent
  condensation models frequently assume that $\tau_s \bar{r}/r$ is
  constant, equal to $\langle \tau_s \rangle$, in order to obtain
  analytical solutions \citep[see
  e.g.][]{sardina2015continuous,field2014mixed}.  This can be
  justified globally, as the assumption of constant
  $\langle R^3\rangle$ implies $\langle SR \rangle = 0$, see
  Eq.~(\ref{eq:ModelGeneralMomentsR}).  Under such approximations the
  evolution of $S$ greatly simplifies: Equation (\ref{eq:ModelS})
  loses its dependence on an integral quantity
  ($\langle S \rangle_E \approx 0$), as well as its nonlinearity
  ($AR \approx \mbox{const.}$).  Hence, the dynamics of $S$ are
  independent of $R$ and follow a standard Ornstein--Uhlenbeck
  process, where $R$ is just the time integration of this
  process. Therefore, the distributions of $S$ and $R$ are Gaussian
  and are fully determined by their first two moments. Using
  (\ref{eq:ModelGeneralMoments}), the distributions of $S$ and $R^2$
  can be straightforwardly written as

\begin{subequations}\label{eq:ModelShortTime}
\begin{align}
S &\sim \mathcal{N} \left( 0 \; , \; \Theta + ( 1 - \Theta  ) \exp\left( - {2 T}/{\Theta} \right) \right) \\
R^2 &\sim \mathcal{N} \left( R_0^2, \Theta^2 \left[ 2T +  (4\Theta-2)\left(\exp\left( - { T}/{\Theta} \right) - 1\right) + (1-\Theta)\left(\exp\left( - {2 T}/{\Theta} \right) - 1\right) \right] \right) \label{eq:ModelR2ShortTime},
\end{align}
\end{subequations}
with
\begin{equation}\label{eq:ShortTimeTimeScaleOmega}
\Theta = \frac{1}{1+AR} = \frac{1}{1+ t_s/\tau_s} = \frac{1/t_s}{1/t_s + 1/\tau_s} = \frac{\tau_s}{\tau_s+ t_s} < 1 ,
\end{equation}
where we have used the assumptions that $\left(1+AR\right)$
  and $\langle \tau_s \rangle$ are constant.

These predictions are compared in Fig.~\ref{fig:ShortTimeBehavior} to
the DNS results for the case $C1$.  The initial time evolution of the
relative standard deviation matches very well.  This demonstrates that
the droplets feel the supersaturation along their trajectory as if it
was white noise because $\Theta \ll \tau_c/t_s$ (see
\S\ref{subsec:HandWavyingArguments}).  The resulting Brownian motion
for the drop surface area leads to a Gaussian distribution with a
variance increasing linearly with time.  The time-dependent DNS
distributions can be perfectly collapsed to a standard normal
distribution using the predicted values.  When time increases, the
assumptions used to derive Eqs.~(\ref{eq:ModelShortTime}) are
violated, \textit{e.g.}, $\langle R^3\rangle$ increases slightly with
increasing $\sigma_{R^2}$.  Nevertheless, the prediction for
$\sigma_{R^2}/\langle R^2\rangle$ stays close to the DNS values.

\begin{figure}
  \centerline{ \includegraphics[width=0.65\textwidth]{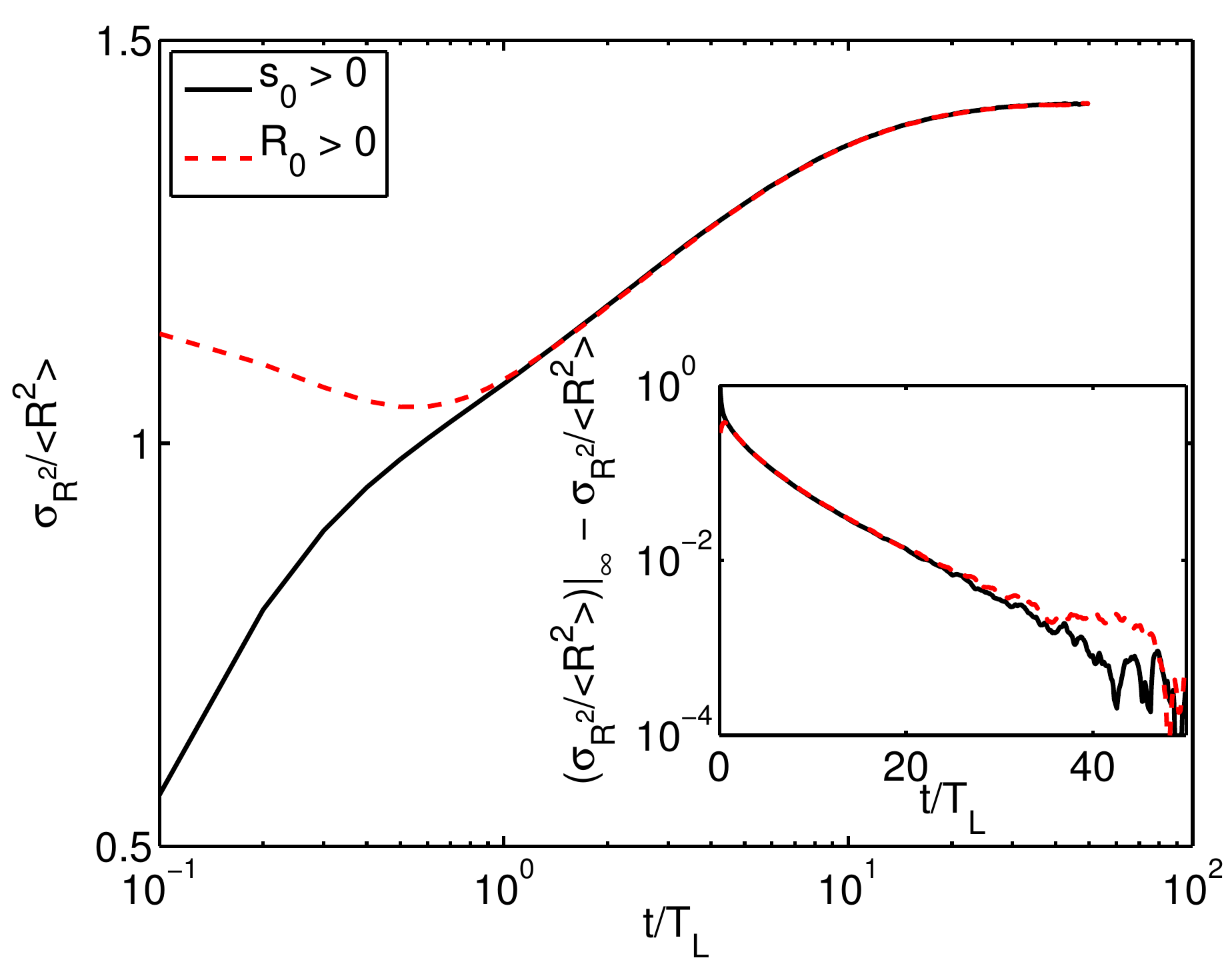}}
  \caption{Time behaviour of the droplet surface normalized standard
    deviation $\sigma_{R^2}/\langle R^2\rangle$ in lin-log for two
    different initial conditions but the same model parameters $A$ and
    $\langle W\rangle$.  Red dashed line: Initially delta function in
    $R>0$ and zero mean normal distribution for $s$; Black solid line:
    Initially zero sized droplets ($R=0$) and constant positive
    $S>0$.  Inset: Same data but subtracted to the steady state value
    and in a log-lin plot.}
\label{fig:convergence_diff_IC}
\end{figure}
The system behaviour changes drastically when the drop size
distribution becomes so wide that some droplets start to completely
evaporate.  
Thereby, they lose their memory of their past history and they thus become independent of their initial condition.
This is apparent in
Fig.~\ref{fig:convergence_diff_IC} where the time evolutions of two
MC simulations with different initial conditions but the same parameters
$A$ and $\langle W\rangle$ are compared.  While the time evolutions
depend on the initial conditions at short times and is thus very different,
they converge towards each other and coincide at long times.

The time at which the diffusive short-time behaviour ends and a
transition to the long-time behaviour starts is the time at which
complete evaporation becomes significant. This evaporation time
$T_{\rm evap}$ can be estimated as the time where
$2\sigma_{R^2} \sim \langle R^2 \rangle$. From (\ref{eq:ModelR2ShortTime}) one obtains (in the limit of $T \gg \Theta$)
\begin{equation}
\label{eq:Tevap}
\frac{t_{\rm evap}}{t_s} \sim \frac{1}{8 \Theta^2}\frac{\tau_c^2}{t_s^2} -  \frac{1}{2} + \frac{3}{2} \Theta .
\end{equation}
This time scale was already shown in
Fig.~\ref{fig:compare_convergence} to mark the transition from the
diffusive behaviour to the convergence to the steady state.  If we
want to observe the diffusive behaviour on timescales of the order of
$t_s$, the time until first evaporation has to satisfy
$t_{\rm evap} > t_s$. For $\Theta$ small, this leads to
\begin{equation}
\tau_c > \frac{\sqrt{8}}{1/t_s + 1/\tau_s} .
\label{eq:TauCForSqrtTBehavoir}
\end{equation}
Note that a harmonic mean of $t_s$ and $\tau_s$ appears ---\,see
Eq.~(\ref{eq:ShortTimeTimeScaleOmega}).  This is very close to the
initial phenomenological reasoning
$\tau_c > \min\left(\tau_s,t_0\right)$ that was used to distinguish
between the two regimes found in the DNS (see
\S\ref{subsec:HandWavyingArguments}).  Also \citet{paoli2009turbulent}
could collapse the time dependence of their DNS results by normalizing
time with the harmonic mean of $t_0$ and $\tau_s$.  As rule of thumb
it can be stated that the short-time behaviour containing no
evaporated droplets is always present.  However, when
$\langle W\rangle$ is smaller than one, it is too short to be
identified.

\subsection{Long-time Behaviour} \label{subsec:LongTimeBehavior}

\begin{figure}
  \centerline{ \includegraphics[width=0.65\textwidth]{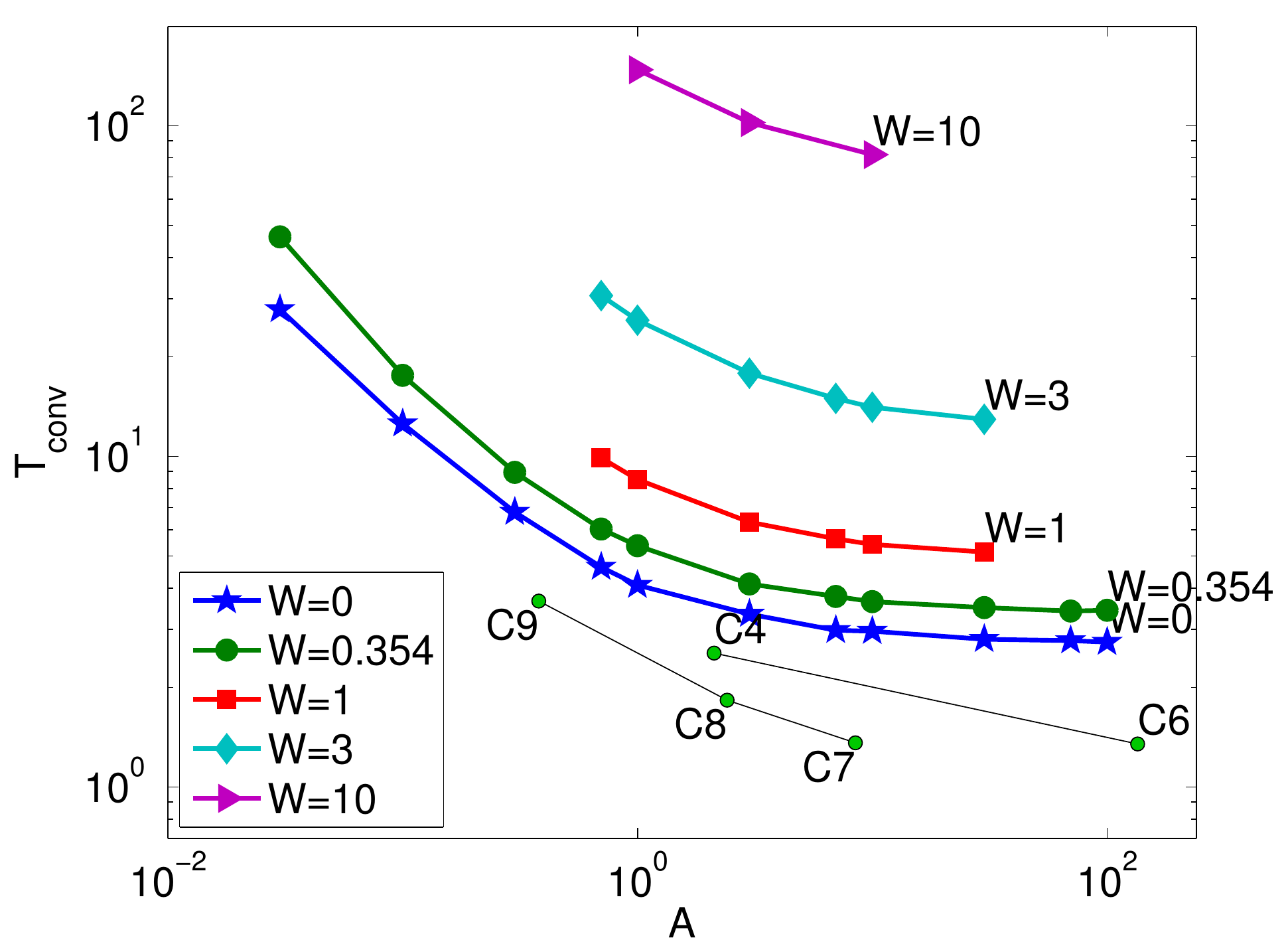}}
  \caption{Exponential convergence time scale $T_{conv}$ as a function
    of the parameter A and with the Scharparameter $\langle W\rangle$
    ranging from $0$ to $10$.  As in fig.~\ref{fig:steady_sate_fn_a}
    the DNS values of the cases $C7$, $C8$, and $C9$ with $\langle
    W\rangle \sim 0$ and $C4$ and $C6$ with $\langle W\rangle \sim
    0.354$ are superimposed.
  }
\label{fig:exponential_rate}
\end{figure}

The long-time evolution of the system is independent of the initial
conditions (see Fig.~\ref{fig:convergence_diff_IC}).  It can be
expected that the system converges exponentially fast to the steady
state with a rate given by the largest eigenvalue of the Fokker-Planck
equation associated with the stochastic dynamics (\ref{eq:Model}). This
is confirmed in the inset of Fig.~\ref{fig:convergence_diff_IC}.  The
exponential convergence time $T_{\rm conv}$ can be measured and is
plotted in Fig.~\ref{fig:exponential_rate}.  The curves are very
similar to the steady state values of $\langle s\rangle_E$ and the
behaviour can be explained by the same arguments.  The larger is the
parameter $A$, the stronger is the bias due to the coupling term,
\textit{i.e.}\/ the waiting time to be evaporated shortens.  For large
$A$ the convergence time seems to saturate to approximately
$2.75\,t_s$ for $\langle W\rangle=0$.  At larger masses
$\langle W\rangle$, the convergence to the steady state is slower.
Since the mean droplet size is larger (see
Fig.~\ref{fig:steady_sate_fn_a}), the time needed for droplets to
completely evaporate increases.  In DNS the convergence to the steady
state is faster than in the model.  This was already visible from
Fig.~\ref{fig:compare_convergence}, albeit it can also be seen that
the measurements are less certain and the fluctuations of the moments
of $R$ are stronger.  Nevertheless, the trends with respect to $A$ and
$\langle W\rangle$ seem to be consistent with the behaviour obtained
from the model.

\section{Summary and Concluding Remarks} \label{sec:conclusion}

We have shown in this study that under some rather broad
  assumptions, the problem of turbulent condensation depends on two
  parameters only and can be modeled and understood by a relatively
  simple stochastic system.  The turbulent supersaturation
  fluctuations, together with the turbulent transport and mixing,
  offer different conditions for the growth of droplets.  This can be
  modeled by a Lagrangian stochastic approach for the joint evolution
  of the squared droplet radius and the supersaturation along its
  trajectory.  The two parameters are fixed by the total amount of
  water and the thermodynamic properties, as well as the Lagrangian
  integral timescale of the turbulent supersaturation.  The model
  reproduces well the time evolution of droplet size distributions
  obtained from direct numerical simulations.

With the help of the model we can reconcile the results of
  the DNS studies in the literature by applying our findings
  to the other studies.  Depending on the initial conditions and the
  considered time, different regimes in the evolution of the system
  can be found.  For initially large drops, the time until the first
  evaporation is long (see Eq.~(\ref{eq:Tevap})). Hence,
    first the supersaturation $S$ converges on a timescale
    $\Theta t_s$ from its initially droplet free variance
    $s_{\mathrm {rms}}^2$ to a smaller variance of size
    $\Theta s_{\mathrm {rms}}^2$ (see Eqs.~(\ref{eq:ModelShortTime})).
    This result can be found equal to the stochastic model
      prediction by \citet{field2014mixed} using the dimensional
      arguments of \citet{lanotte2009cloud}
      $s_{\mathrm{rms}} = a_1 u_{\mathrm{rms}} t_0$. Since it shares
    some similarities with the quasi-equilibrium supersaturation value
    in the case of no turbulence but constant supersaturation source
    \citep{gw13}, it is called quasi-steady value in
    \citet{lanotte2009cloud} and \citet{sardina2015continuous}.
    Hence, at times $t \gg \Theta$ the droplet surface area diffuses
    (\S\ref{subsec:ShortTimeBehavior}).  \citet{sardina2015continuous}
    obtained under the assumption $\tau_s \ll t_0$ an analytical
    expression for $\sigma_{r^2}$ (their eq.~13) which reasonably
    matched their DNS results.  Eq.~(\ref{eq:ModelShortTime}) derived
    in \S\ref{subsec:ShortTimeBehavior} provides an identical
    dependence for $\sigma_{r^2}$ if the difference between $t_0$ and
    $t_s$ is ignored and $s_{\mathrm{rms}} = a_1 u_{\mathrm{rms}} t_0$
    is used again. For smaller $\tau_c/t_0$ complete evaporation
  occurs faster \citep{lanotte2009cloud}, so that $\langle s \rangle_E$ becomes negative
  (compare \citet{celani2007droplet}).  Due to the associated memory
  loss the long-time behaviour characterised in
  \S\ref{subsec:LongTimeBehavior} is observed \citep[as for instance
  in][]{celani2005droplet,celani2008equivalent}.  Eventually, the
  steady state is reached (\S\ref{subsec:TheoResults}); the evolution
  of the droplet spectrum stops due to the fact that the average of
  $s$ conditioned on $r>0$ converges to zero
  \citep[][]{celani2009droplet}.  We found that in this statistical
  steady state, the droplet mass distribution exhibits an exponential
  tail.  To conclude, we argue that the problem just features
  different regimes but is actually independent of the dimensionality,
  the kind of supersaturation forcing, the activation process, and so
  on, so that there is strong evidence that it is \emph{universal}. 

In this study we focused on homogeneous turbulent
  condensation.  However, there is no obvious reason why the model
  (Eqs.~\ref{eq:Model}) should not be applied to inhomogenous cases.
  For example the problem of mixing of sub- and supersaturated regions
  is based on the same governing equations.  DNS show even similar
  trends for the PDF of $r^2$ \citep[fig~3. of][]{kumar2012extreme}:
  The tail has a shape in between a Gaussian and an exponential.  For
  problems with supersaturation sources and sinks such as temperature
  gradients, the mass $\langle W\rangle$ (Eq.~\ref{eq:ModelW}) could
  be made dependent of position or time, respectively.

Nevertheless, high droplet volume loading or strong droplet
  inertia would certainly question the validity of the modeling. We
have indeed assumed that the fluctuations of the supersaturation field
along droplets follows a rather simple dynamics close to diffusion.
However, it is known that particles with significant inertia
tend to cluster in the fronts of an advected scalar
\citep{WetchagarunRiley2010,PhysRevLett.112.234503} leading to very
intermittent Lagrangian statistics that cannot be reproduced by the
model.

In the settings where droplet inertia can be neglected, such simple
model can account for most of the effects of turbulence as the
condensation/evaporation process is reversible and only one-drop
one-time statistics are relevant.  However, for the understanding of
the spatial structure of the supersaturation field or the transition
to a collision-induced growth, two-point statistics become important.
Here, we expect non-trivial effects since turbulent transport can
display anomalous scaling laws. By looking closely at
Fig.~\ref{fig:two_snapshots} it can be seen that the droplet size and
the local supersaturation value are correlated for newly activated,
small droplets while this correlation vanishes for larger droplets
whose size changes much slower.  This means it is
overproportionally likely to find equally sized, small droplets on
isosurfaces of specific supersaturation values, \textit{i.e.}\/ it
seems that a condensational clustering exists in absence of any
droplet inertia.

\medskip

We acknowledge H.~Homann for very useful discussions. This research
has received funding from the French Agence Nationale de la Recherche
(Programme Blanc ANR-12-BS09-011-04). This work was performed using
HPC resources from GENCI-TGCC (Grant 2015-2b6815) and from the 
Mesocenter SIGAMM hosted by the Observatoire de la C\^ote d'Azur.

\appendix
\section{}\label{appendixA}
In the manuscript vectorial quantities are set in bold, dimensional
quanitities in lower case, and dimensionless quantities in upper
case.

\subsection*{List of Symbols}
\begin{tabbing}
\parbox[t]{1.5in}{$A$} \=\parbox[t]{4.5in}{coupling parameter} \\
\parbox[t]{1.5in}{$a_1$} \=\parbox[t]{4.5in}{updraft forcing constant} \\
\parbox[t]{1.5in}{$a_2$} \=\parbox[t]{4.5in}{supersaturation constant} \\
\parbox[t]{1.5in}{$a_3$} \=\parbox[t]{4.5in}{condensation constant} \\
\parbox[t]{1.5in}{$l_0$} \=\parbox[t]{4.5in}{large-scale lengthscale} \\
\parbox[t]{1.5in}{$N$} \=\parbox[t]{4.5in}{number of droplets} \\
\parbox[t]{1.5in}{$N_x^3$} \=\parbox[t]{4.5in}{number of spatial collocation points} \\
\parbox[t]{1.5in}{$n_d$} \=\parbox[t]{4.5in}{droplet number density} \\
\parbox[t]{1.5in}{$p$} \=\parbox[t]{4.5in}{pressure} \\
\parbox[t]{1.5in}{$Re$} \=\parbox[t]{4.5in}{large-scale Reynolds number} \\
\parbox[t]{1.5in}{$R_\lambda$} \=\parbox[t]{4.5in}{Taylor-scale based Reynolds numer} \\
\parbox[t]{1.5in}{$r$} \=\parbox[t]{4.5in}{droplet radius} \\
\parbox[t]{1.5in}{$\bar{r}$} \=\parbox[t]{4.5in}{characteristic, mass-averaged droplet radius} \\
\parbox[t]{1.5in}{$St$} \=\parbox[t]{4.5in}{large scale Stokes number} \\
\parbox[t]{1.5in}{$Sc$} \=\parbox[t]{4.5in}{Schmidt number} \\
\parbox[t]{1.5in}{$s$} \=\parbox[t]{4.5in}{local supersaturation} \\
\parbox[t]{1.5in}{$\langle s \rangle_E$} \=\parbox[t]{4.5in}{mean Eulerian supersaturation} \\
\parbox[t]{1.5in}{$s_{rms}$} \=\parbox[t]{4.5in}{supersaturation standard deviation in absence of droplets} \\
\parbox[t]{1.5in}{$t$} \=\parbox[t]{4.5in}{time} \\
\parbox[t]{1.5in}{$t_0$} \=\parbox[t]{4.5in}{large-scale timescale} \\
\parbox[t]{1.5in}{$t_{conv}$} \=\parbox[t]{4.5in}{convergence timescale} \\
\parbox[t]{1.5in}{$t_{evap}$} \=\parbox[t]{4.5in}{time until first evaporation} \\
\parbox[t]{1.5in}{$t_{u}$} \=\parbox[t]{4.5in}{integral timescale of the Lagrangian velocity autocorrelation} \\
\parbox[t]{1.5in}{$t_{s}$} \=\parbox[t]{4.5in}{integral timescale of the Lagrangian supersaturation autocorrelation}\\
\parbox[t]{1.5in}{$\mathbf{u}$} \=\parbox[t]{4.5in}{gasflow velocity} \\
\parbox[t]{1.5in}{$u_{\rm{rms}}$} \=\parbox[t]{4.5in}{root-mean-square velocity} \\
\parbox[t]{1.5in}{$\mathbf{v}$} \=\parbox[t]{4.5in}{droplet velocity} \\
\parbox[t]{1.5in}{$\langle W \rangle$} \=\parbox[t]{4.5in}{total mass parameter} \\
\parbox[t]{1.5in}{$w$} \=\parbox[t]{4.5in}{total water mass} \\
\parbox[t]{1.5in}{$\mathbf{x}$} \=\parbox[t]{4.5in}{position} \\
\parbox[t]{1.5in}{$\varepsilon$} \=\parbox[t]{4.5in}{mean kinetic energy dissipation rate} \\
\parbox[t]{1.5in}{$\xi$} \=\parbox[t]{4.5in}{standard white noise} \\
\parbox[t]{1.5in}{$\eta$} \=\parbox[t]{4.5in}{Kolmogorov dissipative scale} \\
\parbox[t]{1.5in}{$\Theta$} \=\parbox[t]{4.5in}{time scale ratio} \\
\parbox[t]{1.5in}{$\kappa$} \=\parbox[t]{4.5in}{molecular diffusivity of vapour inside the gas} \\
\parbox[t]{1.5in}{$\nu$} \=\parbox[t]{4.5in}{kinematic viscosity of the gasflow} \\
\parbox[t]{1.5in}{$\rho$} \=\parbox[t]{4.5in}{density} \\
\parbox[t]{1.5in}{$\sigma$} \=\parbox[t]{4.5in}{standard deviation} \\
\parbox[t]{1.5in}{$\tau_c$} \=\parbox[t]{4.5in}{condensation timescale} \\
\parbox[t]{1.5in}{$\tau_d$} \=\parbox[t]{4.5in}{droplet response time} \\
\parbox[t]{1.5in}{$\tau_s$} \=\parbox[t]{4.5in}{supersaturation timescale} \\
\parbox[t]{1.5in}{$\tau_\nu$} \=\parbox[t]{4.5in}{viscous mixing time scale} \\
\parbox[t]{1.5in}{$\tau_\kappa$} \=\parbox[t]{4.5in}{diffusion mixing time scale} \\
\parbox[t]{1.5in}{$\tau_\eta$} \=\parbox[t]{4.5in}{Kolmogorov time} \\
\parbox[t]{1.5in}{$\upsilon$} \=\parbox[t]{4.5in}{volume} \\
\parbox[t]{1.5in}{$\phi$} \=\parbox[t]{4.5in}{forcing} \\
\end{tabbing}

\bibliographystyle{jfm}
\bibliography{siewert}

\begin{thebibliography}{39}
\expandafter\ifx\csname natexlab\endcsname\relax\def\natexlab#1{#1}\fi
\def\au#1{#1} \def\ed#1{#1} \def\yr#1{#1}\def\at#1{#1}\def\jt#1{\textit{#1}}
  \def\bt#1{#1}\def\bvol#1{\textbf{#1}} \def\vol#1{#1} \def\pg#1{#1}
  \def\publ#1{#1}\def\arxiv#1{#1}\def\org#1{#1}\def\st#1{\textit{#1}}

\bibitem[Bartlett \& Jonas(1972)]{bartlett1972dispersion}
{\sc \au{Bartlett, J.T.} \& \au{Jonas, P.R.}} \yr{1972}  \at{On the dispersion
  of the sizes of droplets growing by condensation in turbulent clouds}.
  \jt{Quart. J. Roy. Meteor. Soc}  \bvol{98},  \pg{150--164}.

\bibitem[Bec {\em et~al.\/}(2014)Bec, Homann \&
  Krstulovic]{PhysRevLett.112.234503}
{\sc \au{Bec, J.}, \au{Homann, H.} \& \au{Krstulovic, G.}} \yr{2014}
  \at{Clustering, fronts, and heat transfer in turbulent suspensions of heavy
  particles}.  \jt{Phys. Rev. Lett.}  \bvol{112},  \pg{234503}.

\bibitem[Celani {\em et~al.\/}(2005)Celani, Falkovich, Mazzino \&
  Seminara]{celani2005droplet}
{\sc \au{Celani, A.}, \au{Falkovich, G.}, \au{Mazzino, A.} \& \au{Seminara,
  A.}} \yr{2005}  \at{Droplet condensation in turbulent flows}.  \jt{Europhys.
  Lett.}  \bvol{70}~(6),  \pg{775}.

\bibitem[Celani {\em et~al.\/}(2001)Celani, Lanotte, Mazzino \&
  Vergassola]{celani2001fronts}
{\sc \au{Celani, A.}, \au{Lanotte, A.}, \au{Mazzino, A.} \& \au{Vergassola,
  M.}} \yr{2001}  \at{Fronts in passive scalar turbulence}.  \jt{Phys. Fluids}
  \bvol{13}~(6),  \pg{1768--1783}.

\bibitem[Celani {\em et~al.\/}(2007)Celani, Mazzino, Seminara \&
  Tizzi]{celani2007droplet}
{\sc \au{Celani, A.}, \au{Mazzino, A.}, \au{Seminara, A.} \& \au{Tizzi, M.}}
  \yr{2007}  \at{Droplet condensation in two-dimensional bolgiano turbulence}.
  \jt{J. Turbul.}  \bvol{8},  \pg{N17}.

\bibitem[Celani {\em et~al.\/}(2008)Celani, Mazzino \&
  Tizzi]{celani2008equivalent}
{\sc \au{Celani, A.}, \au{Mazzino, A.} \& \au{Tizzi, M.}} \yr{2008}  \at{The
  equivalent size of cloud condensation nuclei}.  \jt{New J. Phys.}
  \bvol{10}~(7),  \pg{075021}.

\bibitem[Celani {\em et~al.\/}(2009)Celani, Mazzino \&
  Tizzi]{celani2009droplet}
{\sc \au{Celani, A.}, \au{Mazzino, A.} \& \au{Tizzi, M.}} \yr{2009}
  \at{Droplet feedback on vapor in a warm cloud}.  \jt{Int. J. Mod Phys B}
  \bvol{23}~(28n29),  \pg{5434--5443}.

\bibitem[Devenish {\em et~al.\/}(2012)Devenish, Bartello, Brenguier, Collins,
  Grabowski, {IJ}zermans, Malinowski, Reeks, Vassilicos, Wang \& Warhaft]{db12}
{\sc \au{Devenish, B.J.}, \au{Bartello, P.}, \au{Brenguier, J.-L.},
  \au{Collins, L.R.}, \au{Grabowski, W.W.}, \au{{IJ}zermans, R.H.A.},
  \au{Malinowski, S.P.}, \au{Reeks, M.W.}, \au{Vassilicos, J.-C.}, \au{Wang,
  L.-P.} \& \au{Warhaft, Z.}} \yr{2012}  \at{Droplet growth in warm turbulent
  clouds}.  \jt{Q. J. R. Meteorol. Soc.}  \pg{pp. 1401--1429},
  {DOI}:10.1002/qj.1897.

\bibitem[Devenish {\em et~al.\/}(2016)Devenish, Furtado \&
  Thomson]{devenish2016analytical}
{\sc \au{Devenish, BJ}, \au{Furtado, K} \& \au{Thomson, DJ}} \yr{2016}
  \at{Analytical solutions of the supersaturation equation for a warm cloud}.
  \jt{J. Atmos. Sci.}  \bvol{73}~(9),  \pg{3453--3465}.

\bibitem[Field {\em et~al.\/}(2014)Field, Hill, Furtado \&
  Korolev]{field2014mixed}
{\sc \au{Field, PR}, \au{Hill, AA}, \au{Furtado, K} \& \au{Korolev, A}}
  \yr{2014}  \at{Mixed-phase clouds in a turbulent environment. part 2:
  Analytic treatment}.  \jt{Q. J. R. Meteorol. Soc.}  \bvol{140}~(680),
  \pg{870--880}.

\bibitem[Gotoh \& Watanabe(2015)]{PhysRevLett.115.114502}
{\sc \au{Gotoh, T.} \& \au{Watanabe, T.}} \yr{2015}  \at{Power and nonpower
  laws of passive scalar moments convected by isotropic turbulence}.  \jt{Phys.
  Rev. Lett.}  \bvol{115},  \pg{114502}.

\bibitem[Grabowski \& Wang(2013)]{gw13}
{\sc \au{Grabowski, W.W.} \& \au{Wang, L.-P.}} \yr{2013}  \at{Growth of cloud
  droplets in a turbulent environment}.  \jt{Annu. Rev. Fluid Mech.}
  \bvol{45}~(1),  \pg{293--324}.

\bibitem[Homann {\em et~al.\/}(2007)Homann, Dreher \& Grauer]{homann2007impact}
{\sc \au{Homann, H.}, \au{Dreher, J.} \& \au{Grauer, R.}} \yr{2007}  \at{Impact
  of the floating-point precision and interpolation scheme on the results of
  dns of turbulence by pseudo-spectral codes}.  \jt{Comp. Phys. Comm.}
  \bvol{177}~(7),  \pg{560--565}.

\bibitem[Ingersoll {\em et~al.\/}(2004)Ingersoll, Dowling, Gierasch, Orton,
  Read, S{\'a}nchez-Lavega, Showman, Simon-Miller \&
  Vasavada]{ingersoll2004dynamics}
{\sc \au{Ingersoll, Andrew~P}, \au{Dowling, Timothy~E}, \au{Gierasch, Peter~J},
  \au{Orton, Glenn~S}, \au{Read, Peter~L}, \au{S{\'a}nchez-Lavega, Agustin},
  \au{Showman, Adam~P}, \au{Simon-Miller, Amy~A} \& \au{Vasavada, Ashwin~R}}
  \yr{2004}  \at{Dynamics of jupiter’s atmosphere}.  \bt{In {\em Jupiter: The
  Planet, Satellites and Magnetosphere\/} (ed. \ed{F.~Bagenal, T.E. Dowling \&
  W.B. McKinnon})}, chap.~6.  \publ{Cambridge Univ. Press}.

\bibitem[Khvorostyanov \& Curry(1999)]{khvorostyanov1999toward}
{\sc \au{Khvorostyanov, V.I.} \& \au{Curry, J.A.}} \yr{1999}  \at{Toward the
  theory of stochastic condensation in clouds. {P}art {I}: A general kinetic
  equation}.  \jt{J. Atmos. Sci.}  \bvol{56}~(23),  \pg{3985--3996}.

\bibitem[Kulmala {\em et~al.\/}(1997)Kulmala, Rannik, Zapadinsky \&
  Clement]{kulmala1997effect}
{\sc \au{Kulmala, M.}, \au{Rannik, {\"U}.}, \au{Zapadinsky, E.L.} \&
  \au{Clement, C.F.}} \yr{1997}  \at{The effect of saturation fluctuations on
  droplet growth}.  \jt{J. Atmos. Sci.}  \bvol{28}~(8),  \pg{1395--1409}.

\bibitem[Kumar {\em et~al.\/}(2012)Kumar, Janetzko, Schumacher \&
  Shaw]{kumar2012extreme}
{\sc \au{Kumar, B.}, \au{Janetzko, F.}, \au{Schumacher, J.} \& \au{Shaw, R.A.}}
  \yr{2012}  \at{Extreme responses of a coupled scalar--particle system during
  turbulent mixing}.  \jt{New J. Phys.}  \bvol{14}~(11),  \pg{115020}.

\bibitem[Lanotte {\em et~al.\/}(2009)Lanotte, Seminara \&
  Toschi]{lanotte2009cloud}
{\sc \au{Lanotte, A.S.}, \au{Seminara, A.} \& \au{Toschi, F.}} \yr{2009}
  \at{Cloud droplet growth by condensation in homogeneous isotropic
  turbulence}.  \jt{J. Atmos. Sci.}  \bvol{66}~(6),  \pg{1685--1697}.

\bibitem[Lasher-Trapp {\em et~al.\/}(2005)Lasher-Trapp, Cooper \&
  Blyth]{lasher2005broadening}
{\sc \au{Lasher-Trapp, S.G.}, \au{Cooper, W.A.} \& \au{Blyth, A.M.}} \yr{2005}
  \at{Broadening of droplet size distributions from entrainment and mixing in a
  cumulus cloud}.  \jt{Q. J. R. Meteorol. Soc.}  \bvol{131}~(605),
  \pg{195--220}.

\bibitem[Lehmann {\em et~al.\/}(2009)Lehmann, Siebert \&
  Shaw]{lehmann2009homogeneous}
{\sc \au{Lehmann, K.}, \au{Siebert, H.} \& \au{Shaw, R.A.}} \yr{2009}
  \at{Homogeneous and inhomogeneous mixing in cumulus clouds: Dependence on
  local turbulence structure}.  \jt{J. Atmos. Sci.}  \bvol{66}~(12),
  \pg{3641--3659}.

\bibitem[McGraw \& Liu(2006)]{mcgraw2006brownian}
{\sc \au{McGraw, R.} \& \au{Liu, Y.}} \yr{2006}  \at{Brownian drift-diffusion
  model for evolution of droplet size distributions in turbulent clouds}.
  \jt{Geophys. Res. Lett.}  \bvol{33}~(3).

\bibitem[Paoli \& Shariff(2009)]{paoli2009turbulent}
{\sc \au{Paoli, R.} \& \au{Shariff, K.}} \yr{2009}  \at{Turbulent condensation
  of droplets: {D}irect simulation and a stochastic model}.  \jt{J. Atmos.
  Sci.}  \bvol{66}~(3),  \pg{723--740}.

\bibitem[Pinsky {\em et~al.\/}(2013)Pinsky, Mazin, Korolev \&
  Khain]{pinsky2013supersaturation}
{\sc \au{Pinsky, M}, \au{Mazin, IP}, \au{Korolev, A} \& \au{Khain, A}}
  \yr{2013}  \at{Supersaturation and diffusional droplet growth in liquid
  clouds}.  \jt{J. Atmos. Sci.}  \bvol{70}~(9),  \pg{2778--2793}.

\bibitem[Pope(2000)]{p00}
{\sc \au{Pope, S.~B.}} \yr{2000} {\em Turbulent flows\/}.  \publ{Cambridge:
  Cambridge University Press}.

\bibitem[Pruppacher \& Klett(1997)]{pk97b}
{\sc \au{Pruppacher, H.} \& \au{Klett, J.}} \yr{1997} {\em Microphysics of
  clouds and precipitation\/}.  \publ{Kluwer Academic Publishers}.

\bibitem[Reveillon \& Demoulin(2007)]{reveillon2007effects}
{\sc \au{Reveillon, J.} \& \au{Demoulin, F.-X.}} \yr{2007}  \at{Effects of the
  preferential segregation of droplets on evaporation and turbulent mixing}.
  \jt{J. Fluid Mech.}  \bvol{583},  \pg{273--302}.

\bibitem[Sancho {\em et~al.\/}(1982)Sancho, San~Miguel \&
  D{\"u}rr]{sancho1982adiabatic}
{\sc \au{Sancho, J.M.}, \au{San~Miguel, M.} \& \au{D{\"u}rr, D.}} \yr{1982}
  \at{Adiabatic elimination for systems of brownian particles with nonconstant
  damping coefficients}.  \jt{J. Stat. Phys.}  \bvol{28}~(2),  \pg{291--305}.

\bibitem[Sardina {\em et~al.\/}(2015)Sardina, Picano, Brandt \&
  Caballero]{sardina2015continuous}
{\sc \au{Sardina, G.}, \au{Picano, F.}, \au{Brandt, L.} \& \au{Caballero, R.}}
  \yr{2015}  \at{Continuous growth of droplet size variance due to condensation
  in turbulent clouds}.  \jt{Phys. Rev. Lett.}  \bvol{115},  \pg{184501}.

\bibitem[Seifert \& Beheng(2006)]{seifert2006two}
{\sc \au{Seifert, A} \& \au{Beheng, KD}} \yr{2006}  \at{A two-moment cloud
  microphysics parameterization for mixed-phase clouds. {P}art 1: Model
  description}.  \jt{Meteorol. Atmos. Phys.}  \bvol{92}~(1-2),  \pg{45--66}.

\bibitem[Shaw(2003)]{s03}
{\sc \au{Shaw, R.~A.}} \yr{2003}  \at{Particle-turbulence interactions in
  atmospheric clouds}.  \jt{Annu. Rev. Fluid Mech.}  \bvol{35},  \pg{183--227}.

\bibitem[Sidin {\em et~al.\/}(2009)Sidin, {IJ}zermans \&
  Reeks]{sidin2009lagrangian}
{\sc \au{Sidin, R.S.}, \au{{IJ}zermans, R.H.} \& \au{Reeks, M.W.}} \yr{2009}
  \at{A {L}agrangian approach to droplet condensation in atmospheric clouds}.
  \jt{Phys. Fluids}  \bvol{21}~(10),  \pg{106603}.

\bibitem[Squires(1952)]{squires1952growth}
{\sc \au{Squires, P}} \yr{1952}  \at{The growth of cloud drops by condensation.
  {I}. {G}eneral characteristics}.  \jt{J. Sci. Res. A}  \bvol{5}~(1),
  \pg{59--86}.

\bibitem[Srivastava(1989)]{srivastava1989growth}
{\sc \au{Srivastava, R.C.}} \yr{1989}  \at{Growth of cloud drops by
  condensation: A criticism of currently accepted theory and a new approach}.
  \jt{J. Atmos. Sci.}  \bvol{46}~(7),  \pg{869--887}.

\bibitem[Twomey(1959)]{twomey1959nuclei}
{\sc \au{Twomey, S}} \yr{1959}  \at{The nuclei of natural cloud formation part
  {II}: The supersaturation in natural clouds and the variation of cloud
  droplet concentration}.  \jt{Geofisica pura e applicata}  \bvol{43}~(1),
  \pg{243--249}.

\bibitem[Vaillancourt {\em et~al.\/}(2002)Vaillancourt, Yau, Bartello \&
  Grabowski]{vaillancourt2002microscopic}
{\sc \au{Vaillancourt, P.A.}, \au{Yau, M.K.}, \au{Bartello, P.} \&
  \au{Grabowski, W.W.}} \yr{2002}  \at{Microscopic approach to cloud droplet
  growth by condensation. {P}art {II}: Turbulence, clustering, and
  condensational growth}.  \jt{J. Atmos. Sci.}  \bvol{59}~(24),
  \pg{3421--3435}.

\bibitem[Vaillancourt {\em et~al.\/}(2001)Vaillancourt, Yau \&
  Grabowski]{vaillancourt2001microscopic}
{\sc \au{Vaillancourt, P.A.}, \au{Yau, M.K.} \& \au{Grabowski, W.W.}} \yr{2001}
   \at{Microscopic approach to cloud droplet growth by condensation. {P}art
  {I}: Model description and results without turbulence}.  \jt{J. Atmos. Sci.}
  \bvol{58}~(14),  \pg{1945--1964}.

\bibitem[Warhaft(2000)]{warhaft2000passive}
{\sc \au{Warhaft, Z}} \yr{2000}  \at{Passive scalars in turbulent flows}.
  \jt{Annu. Rev. Fluid Mech.}  \bvol{32}~(1),  \pg{203--240}.

\bibitem[Wetchagarun \& Riley(2010)]{WetchagarunRiley2010}
{\sc \au{Wetchagarun, S.} \& \au{Riley, J.J.}} \yr{2010}  \at{Dispersion and
  temperature statistics of inertial particles in isotropic turbulence}.
  \jt{Phys. Fluids}  \bvol{22}~(6),  \pg{063301}.

\bibitem[Yeung(2001)]{yeung2001lagrangian}
{\sc \au{Yeung, PK}} \yr{2001}  \at{Lagrangian characteristics of turbulence
  and scalar transport in direct numerical simulations}.  \jt{J. Fluid Mech.}
  \bvol{427},  \pg{241--274}.

\end{thebibliography}

\end{document}